\documentclass[12pt]{article}
\usepackage{epsf,rotating,graphicx,float}
\usepackage{hyperref,supertabular}
\usepackage{amsmath, amsthm, amssymb}
\renewcommand{\thefootnote}{\fnsymbol{footnote}}
\newlength{\pubnumber} 

\catcode`\@=11
\@addtoreset{equation}{section}

\def\section{\@startsection{section}{1}{\z@}{3.5ex plus 1ex minus .2ex}
 {2.3ex plus .2ex}{\large\bf}}
\def\subsection{\@startsection{subsection}{2}{\z@}{2.3ex plus .2ex}
 {2.3ex plus .2ex}{\bf}}

\begin{document} 
\begin{titlepage}
\samepage{
\setcounter{page}{1}
\rightline{LTH--896}
\rightline{BU-HEPP 11-01}
\rightline{CASPER 11-01}
\rightline{\tt hep-ph/}
\vfill
\begin{center}
 {\Large \bf 
Investigation of Quasi--Realistic Heterotic \\String Models with Reduced Higgs Spectrum
}
\vfill
 \vfill
 {\large  G.\  Cleaver\footnote{Gerald{\underline{\phantom{B}}}Cleaver@baylor.edu}$^1$,
A.\  Faraggi\footnote{Faraggi@amtp.liv.ac.uk}$^2$,
J. Greenwald$\footnote{Jared\underline{\phantom{a}}Greenwald@baylor.edu} ^{1}$,
D. Moore$\footnote{Douglas\underline{\phantom{a}}Moore1@baylor.edu}^{1}$,
K. Pechan $\footnote{Kristen.Pechan@physics.tamu.edu}^{3}$,
E. Remkus$\footnote{Erik\underline{\phantom{a}}Remkus@baylor.edu}^{1}$
and
T. Renner $\footnote{Timothy\underline{\phantom{a}}Renner@baylor.edu}^{1}$

}
\vspace{.12in}
 {\it $^{1}$ CASPER, Department of Physics, Baylor University,
            Waco, TX, 76798-7316}\\
{\it  $^{2}$ Department of Mathematical Sciences, University of Liverpool,
                Liverpool L69 7ZL}\\
\vspace{.025in}
{\it $^{3}$ Mitchell Institute, Dept.\  of Physics, Texas A\&M,
            College Station, TX, 77843-4242}\\
\end{center}
\vfill
\begin{abstract}
Quasi--realistic heterotic--string models in the free fermionic formulation typically
contain an anomalous $U(1)$, which gives rise to a Fayet--Iliopolous term that breaks 
supersymmetry at the one--loop level in string perturbation theory. Supersymmetry
is restored by imposing F-- and D--flatness on the vacuum. In \cite{cfmt} we presented a 
three generation free fermionic standard--like model which did not admit stringent 
F-- and D--flat directions, and argued that the all the moduli in the model are fixed. 
The particular property of the model was the reduction of the untwisted Higgs 
spectrum by a combination of symmetric and asymmetric boundary conditions 
with respect to the internal fermions associated with the compactified dimensions. 
In this paper we extend the analysis of free fermionic models with reduced 
Higgs spectrum to the cases in which the $SO(10)$ symmetry is left unbroken, 
or is reduced to the flipped $SU(5)$ subgroup. 
We show that all the models that we study in this paper do admit stringent 
flat directions. The only examples of models that do not admit stringent flat directions 
remain the strandard--like models of ref. \cite{cfmt}. 
\end{abstract}
\smallskip}
\end{titlepage}

\renewcommand{\thefootnote}{\arabic{footnote}}
\setcounter{footnote}{0}

\def\l{\label}
\def\beq{\begin{equation}}
\def\eeq{\end{equation}}
\def\beqn{\begin{eqnarray}}
\def\eeqn{\end{eqnarray}}
\def\nolabel{\nonumber }

\def\ie{{\it i.e.}}
\def\eg{{\it e.g.}}
\def\half{{\textstyle{1\over 2}}}
\def\third{{\textstyle {1\over3}}}
\def\quarter{{\textstyle {1\over4}}}
\def\tenth{{\textstyle {1\over{10}}}}
\def\m{{\tt -}}
\def\p{{\tt +}}

\def\slash#1{#1\hskip-6pt/\hskip6pt}
\def\slk{\slash{k}}
\def\GeV{\,{\rm GeV}}
\def\TeV{\,{\rm TeV}}
\def\y{\,{\rm y}}
\def\SM{Standard-Model }
\def\SUSY{supersymmetry }
\def\SSSM{supersymmetric standard model}
\def\vev#1{\left\langle #1\right\rangle}
\def\l{\langle}
\def\r{\rangle}

\def\Htw{{\tilde H}}
\def\chibar{{\overline{\chi}}}
\def\qbar{{\overline{q}}}
\def\ibar{{\overline{\imath}}}
\def\jbar{{\overline{\jmath}}}
\def\Hbar{{\overline{H}}}
\def\Qbar{{\overline{Q}}}
\def\abar{{\overline{a}}}
\def\alphabar{{\overline{\alpha}}}
\def\betabar{{\overline{\beta}}}
\def\tautwo{{ \tau_2 }}
\def\thetatwo{{ \vartheta_2 }}
\def\thetathree{{ \vartheta_3 }}
\def\thetafour{{ \vartheta_4 }}
\def\ttwo{{\vartheta_2}}
\def\tthree{{\vartheta_3}}
\def\tfour{{\vartheta_4}}
\def\ti{{\vartheta_i}}
\def\tj{{\vartheta_j}}
\def\tk{{\vartheta_k}}
\def\calF{{\cal F}}
\def\smallmatrix#1#2#3#4{{ {{#1}~{#2}\choose{#3}~{#4}} }}
\def\ab{{\alpha\beta}}
\def\Minv{{ (M^{-1}_\ab)_{ij} }}
\def\bone{{\bf 1}}
\def\ii{{(i)}}
\def\V{{\bf V}}
\def\b{{\bf b}}
\def\N{{\bf N}}
\def\t#1#2{{ \Theta\left\lbrack \matrix{ {#1}\cr {#2}\cr }\right\rbrack }}
\def\C#1#2{{ C\left\lbrack \matrix{ {#1}\cr {#2}\cr }\right\rbrack }}
\def\tp#1#2{{ \Theta'\left\lbrack \matrix{ {#1}\cr {#2}\cr }\right\rbrack }}
\def\tpp#1#2{{ \Theta''\left\lbrack \matrix{ {#1}\cr {#2}\cr }\right\rbrack }}
\def\l{\langle}
\def\r{\rangle}


\def\inbar{\,\vrule height1.5ex width.4pt depth0pt}

\def\IC{\relax\hbox{$\inbar\kern-.3em{\rm C}$}}
\def\IQ{\relax\hbox{$\inbar\kern-.3em{\rm Q}$}}
\def\IR{\relax{\rm I\kern-.18em R}}
 \font\cmss=cmss10 \font\cmsss=cmss10 at 7pt
\def\IZ{\relax\ifmmode\mathchoice
 {\hbox{\cmss Z\kern-.4em Z}}{\hbox{\cmss Z\kern-.4em Z}}
 {\lower.9pt\hbox{\cmsss Z\kern-.4em Z}}
 {\lower1.2pt\hbox{\cmsss Z\kern-.4em Z}}\else{\cmss Z\kern-.4em Z}\fi}

\def\AEF{A. Faraggi}
\def\NPB#1#2#3{{Nucl.\ Phys.}\/ {B \bf #1} (#2) #3}
\def\PLB#1#2#3{{Phys.\ Lett.}\/ {B \bf #1} (#2) #3}
\def\PRD#1#2#3{{Phys.\ Rev.}\/ {D \bf #1} (#2) #3}
\def\PRL#1#2#3{{Phys.\ Rev.\ Lett.}\/ {\bf #1} (#2) #3}
\def\PRP#1#2#3{{Phys.\ Rep.}\/ {\bf#1} (#2) #3}
\def\MODA#1#2#3{{Mod.\ Phys.\ Lett.}\/ {\bf A#1} (#2) #3}
\def\IJMP#1#2#3{{Int.\ J.\ Mod.\ Phys.}\/ {A \bf #1} (#2) #3}
\def\nuvc#1#2#3{{Nuovo Cimento}\/ {\bf #1A} (#2) #3}
\def\JHEP#1#2#3{{JHEP} {\textbf #1}, (#2) #3}
\def\EPJC#1#2#3{{Eur.\ Phys.\ Jour.}\/ {C \bf #1} (#2) #3}
\def\etal{{\it et al\/}}

\newcommand{\be}{\begin{equation}}
\newcommand{\ee}{\end{equation}}
\newcommand{\ba}{\begin{eqnarray}}
\newcommand{\ea}{\end{eqnarray}}
\hyphenation{su-per-sym-met-ric non-su-per-sym-met-ric}
\hyphenation{space-time-super-sym-met-ric}
\hyphenation{mod-u-lar mod-u-lar--in-var-i-ant}
\def\bda#1{{${\cal{D}}_{#1}$}}
\topmargin=20mm
\setcounter{footnote}{0}
\section{Introduction}
\bigskip

String theory provides a viable framework to probe the unification
of gravity and the gauge interactions. Progress in this endeavor 
mandates the development of detailed phenomenological models
as well as improved understanding of the mathematical structures 
that underly the theory. 
On the other hand the Standard Particle Model is compatible 
with all contemporary terrestrial and extra--terrestrial
experimental data. Furthermore, the particle physics 
data is compatible with the hypothesis that the perturbative 
Standard Model remains unaltered up to the Planck scale, 
and that the particle spectrum is embedded in representations 
of a Grand Unified Theory (GUT). Most appealing in this
context is $SO(10)$ unification in which each Standard Model (SM)
generation is embedded in a single spinorial {\bf 16} representation.

Heterotic string theory naturally produces models that
preserve the grand unified embedding of the Standard Model 
states \cite{heteroticstringpheno}.
Among the most realistic string models 
\cite{fsu5,fny,alr,eu,top,cfn,cfs,exophobic}
constructed to date are
the heterotic--string models in the free fermionic formulation,
which are related to $\mathbb{Z}_2\times {\mathbb Z}_2$ toroidal orbifold 
compactifications \cite{foc}. In these models the $SO(10)$ GUT symmetry is 
broken directly at the string level, and include: 
the flipped $SU(5)$ string models \cite{fsu5};
the SM-like string models \cite{fny,eu, cfn}; 
the Pati--Salam string models \cite{alr, exophobic};
and the Left--Right symmetric string models \cite{cfs} (LRS). 
Generically, string vacua contain numerous fields beyond the 
SM states and these must be given sufficiently
heavy mass. Heterotic string models necessarily also 
contain exotic fractionally charged states that are severely constrained
by observations. These can be given mass on the order of the
Planck scale by Standard Model singlet VEVs \cite{cfn}, 
or may be entirely absent from the massless spectrum \cite{exophobic}.
The free fermionic heterotic string models produced
the first known string models in which the matter content of
the effective low energy observable sector consists solely
of the Minimal Supersymmetric Standard Model (MSSM) \cite{cfn}.
The issue of the supersymmetric moduli space and supersymmetry
breaking remains one of the important open issues in 
string phenomenology.  

The relation of free fermion models to ${\mathbb Z}_2\times {\mathbb Z}_2$ orbifolds
mandates the production of three untwisted pairs of vectorial $10$
representations, which decompose as $\left[{\bf 5}\oplus {{\bf \bar 5}}\right]$ 
under $SU(5)$ and as 
$\left[({{\bf  \bar  3}},{\bf 1})_{1/3}+({\bf 1},{\bf 2})_{-1/2}\right]
\oplus
\left[({{\bf 3}},{\bf 1})_{-1/3}+({\bf 1},{\bf 2})_{1/2}\right]$
under the Standard Model $SU(3)_C\times SU(2)_L\times U(1)_Y$ 
gauge group. 
This presents an excess beyond the two electroweak Higgs
doublets that are required in the MSSM. In the Pati--Salam
and Standard--like string models either the color triplets 
or electroweak Higgs doublets from each pair may be projected 
from the massless spectrum by asymmetric/symmetric 
boundary conditions with respect to a given twisted
plane \cite{ps}. Assignment of symmetric and asymmetric 
boundary conditions in two different basis vectors results
in the projection of both color triplets
and electroweak doublets and therefore reduces 
the spectrum without the need to give them heavy mass
by singlet VEVs \cite{fmt,cfmt}. 
In this manner entire {\bf 10} vectorial pairs can be projected.    
An additional consequence of the symmetric plus asymmetric 
assignment is that untwisted $SO(10)$ singlet states are projected
as well and the supersymmetric moduli space is 
constrained \cite{fmt,cfmt}. 

Standard--like string models that employ the symmetric plus 
asymmetric boundary condition assignment to reduce the light
Higgs spectrum were constructed in refs. \cite{fmt,cfmt}. 
The model constructed in ref. \cite{fmt} did not admit flat 
directions that preserve the Standard Model (SM) gauge group.
This gave rise to the possibility that the model does
not admit supersymmetric flat directions at all. 
This would be a very interesting situation as it would 
indicate that supersymmetry is broken perturbatively in
such models. In ref. \cite{cfmt} this question was examined 
further by analyzing the so--called stringent flat directions. 
It was demonstrated that there exists 
SM--like string models with reduced untwisted Higgs
spectrum, which do not admit stringent flat directions 
to any order of non-renomalizable superpotential term. 
This outcome is in contrast to the case of all other
quasi--realistic free fermionic models, which have
been shown to admit such stringent flat directions. 

\begin{figure}[!h]
\begin{center}
\includegraphics[width=4in]{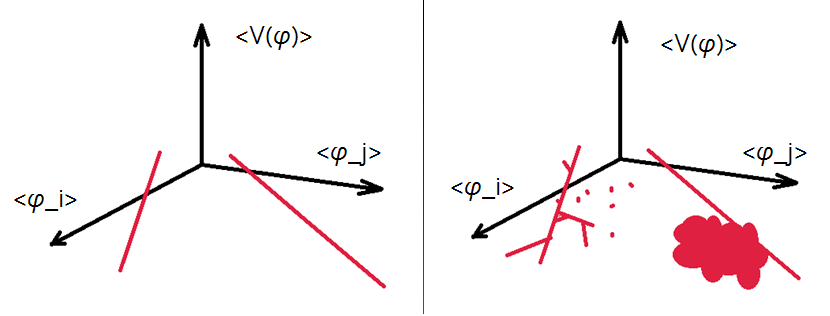}
\end{center}	
\caption{On the left, one will find general characteristics of a model by investigating the stringent flat directions (`root' directions). It seems that every all-order flat direction (right) is attached to or near a `root' direction found via the stringency requirement.}
\label{stringentfig}
\end{figure}

The question therefore arises whether the absence of 
stringent flat directions (see Figure \ref{stringentfig}) in free fermion heterotic--string
models with reduced Higgs spectrum is particular to the 
SM--like models, and whether similar results arise 
in cases in which the $SO(10)$ subgroup is broken to an alternative
subgroup. In this paper we examine this question by analyzing the 
$D$- and $F$--flat directions in four models. Three in which 
the $SO(10)$ symmetry is broken to the the flipped $SU(5)$ 
subgroup and one in which the $SO(10)$ symmetry remains unbroken.
We show that all of these cases admit stringent supersymmetric flat 
directions. Hence, SM--like models 
with reduced Higgs spectrum remain the only examples
in which stringent flat directions have been shown not to exist. 

Our paper is organized as follows: Section \ref{review} contains a review of Free Fermionic Models construction using the NAHE set and additional boundary vectors to control gauge and SUSY breaking. In Section \ref{flatdirection}, we review the process of flat directions used with EFTs as pertaining to Free Fermionic strings.  Here there is a further discussion on stringent $F$-flat directions.  The string models in question are reviewed in Sections \ref{stringmodelsection} and \ref{flatdir}, where we list the boundary vectors pertaining to each model as well as summarize the flat directions.  We conclude our discussion in Section \ref{conclude} and make comment on the impact of our results on future studies of the Free Fermionic Heterotic Landscape.  

\section{Free Fermionic Models}\label{review}

In this section we briefly review the construction and structure of the
free fermionic standard like models.
The notation and further details of the construction of these
models are given elsewhere \cite{fny,eu,nahe,cfn,cfs,fmt}.
In the free fermionic formulation of the heterotic string
in four dimensions \cite{fff} all the world--sheet
degrees of freedom,  required to cancel
the conformal anomaly, are represented in terms of free fermions
propagating on the string world--sheet.
In the light--cone gauge the world--sheet field content consists
of two transverse left-- and right--moving space--time coordinate bosons,
$X_{1,2}^\mu$ and ${\bar X}_{1,2}^\mu$,
and their left--moving fermionic superpartners, $\psi^\mu_{1,2}$,
and additional 62 purely internal
Majorana--Weyl fermions, of which 18 are left--moving
and 44 are right--moving.
The models are constructed by specifying the phases produced by
the world--sheet fermions when transported along the torus'
non--contractible loops
\be  f \rightarrow -e^{i\pi\alpha(f)}f, \quad \alpha(f) \in (-1,1] \label{alphaeq}.\ee
Each model corresponds to a particular choice
of fermion phases consistent with modular invariance and is generated
by a set of basis vectors describing the transformation properties
of the 64 world--sheet fermions.
The physical spectrum is obtained by applying the generalized GSO projections. The GSO projection is administered through the following equations:
\beqn
&&{\bf V}_j \cdot {\bf Q_\alpha} = \left( \sum_i k_{i,j}a_i \right) + s_j \ \ \ (mod \ 2) \\
&&{\boldsymbol \alpha }(f) = \sum_{i=1}^{D} a_i {\bf V}_i \ \ (mod \ 2), \ \  \alpha_i \in \{ {\mathbb{Z}} | 0 \le \alpha \le N_i-1 \}\\
&&{\bf Q}(f) \equiv \frac{1}{2}{\boldsymbol \alpha}(f) + {\bf F}(f)
\eeqn
where ${\bf \alpha}(f)$ is the boundary vector defined by eq. \ref{alphaeq} and is expanded in terms of basis vectors, ${\bf V}$.  For all 64 components of ${\bf V}_i$, $N_i$ is the smallest positive integer such that $N_i{\bf V}_i=0 \ \ (mod \ 2)$.  ${{\bf F}}$ is the fermion number operator and is equal to $\{0,\pm 1\}$ for non-periodic fermions and $\{0,-1\}$ for periodic.  After construction of the string model, the low energy effective field theory is obtained by S--matrix elements between external states \cite{kln}.

The boundary condition basis defining a typical
realistic free fermionic heterotic string model is
constructed in two stages.
The first stage consists of the NAHE set,
which is a set of five boundary condition basis vectors,
$\{{\bf1},S,b_1,b_2,b_3\}$ \cite{nahe}.
The gauge group, after imposing the GSO projections induced
by the NAHE set, is $SO(10)\times SO(6)^3\times E_8$,
with $N=1$ supersymmetry.
The NAHE set divides the internal world--sheet
fermions in the following way: ${\bar\phi}^{1,\cdots,8}$ generate the
hidden $E_8$ gauge group, ${\bar\psi}^{1,\cdots,5}$ generate the $SO(10)$
gauge group, while $\{{\bar y}^{3,\cdots,6},{\bar\eta}^1\}$,
$\{{\bar y}^1,{\bar y}^2,{\bar\omega}^5,{\bar\omega}^6,{\bar\eta}^2\}$ and 
$\{{\bar\omega}^{1,\cdots,4},{\bar\eta}^3\}$ generate the three horizontal
$SO(6)$ symmetries. The left--moving $\{y,\omega\}$ states are divided
into $\{{y}^{3,\cdots,6}\}$,
$\{{y}^1,{y}^2,{\omega}^5,{\omega}^6\}$,
$\{{\omega}^{1,\cdots,4}\}$, while $\chi^{12}$, $\chi^{34}$, $\chi^{56}$
generate the left--moving $N=2$ world--sheet supersymmetry.

The second stage of the
basis construction consists of adding to the
NAHE set three additional boundary condition basis vectors.
These additional basis vectors reduce the number of generations
to three chiral generations, one from each of the sectors $b_1$,
$b_2$ and $b_3$, and simultaneously break the four dimensional
gauge group. The assignment of boundary conditions to
$\{{\bar\psi}^{1,\cdots,5}\}$ breaks $SO(10)$ to one of its subgroups.
Similarly, the hidden $E_8$ symmetry is broken to one of its
subgroups. The flavor $SO(6)^3$ symmetries in the NAHE--based models
are always broken to flavor $U(1)$ symmetries, as the breaking
of these symmetries is correlated with the number of chiral
generations. Three such $U(1)_j$ symmetries are always obtained
in the NAHE based free fermionic models from the subgroup
of the observable $E_8$, which is orthogonal to $SO(10)$.
These are produced by the world--sheet currents ${\bar\eta}^j{\bar\eta}^{j^*}$
($j=1,2,3$), which are part of the Cartan sub--algebra of the
observable $E_8$. Additional unbroken $U(1)$ symmetries, denoted
typically by $U(1)_j$ ($j=4,5,...$), arise by pairing two real
fermions from the sets
$\{{\bar y}^{3,\cdots,6}\}$,
$\{{\bar y}^{1,2},{\bar\omega}^{5,6}\}$ and
$\{{\bar\omega}^{1,\cdots,4}\}$.
The final observable gauge
group depends on the number of such pairings.
Alternatively, a left--moving real fermion from the sets
$\{{ y}^{3,\cdots,6}\}$, $\{{ y}^{1,2},{\omega}^{5,6}\}$ and
$\{{\omega}^{1,\cdots,4}\}$ may be paired with its respective
right--moving real fermion to form an Ising model operator,
in which case the rank of the right--moving gauge group is reduced by one.
The reduction of untwisted
electroweak Higgs doublets crucially depends on the pairings
of the left-- and right--moving fermions from the set
$\{y,\omega|{\bar y},{\bar\omega}\}^{1\cdots6}$.

Subsequent to constructing the basis vectors and extracting the massless
spectrum, the analysis of the free fermionic models proceeds by
calculating the superpotential. The cubic and higher-order terms in
the superpotential are obtained by evaluating the correlators
\beq
A_N\sim \langle V_1^fV_2^fV_3^b\cdots V_N^b\rangle,
\label{supterms}
\eeq
where $V_i^f$ $(V_i^b)$ are the fermionic (scalar) components
of the vertex operators, using the rules given in~\cite{kln}.

\section{Flat Directions of Free Fermions}\label{flatdirection}
A common feature of many of the quasi--realistic free fermionic 
heterotic--string models is the existence of ``anomalous'' $U(1)$'s generated from compactification or from gauge group breaking \cite{Cleaver:1997rk}.  If multiple $U(1)$'s are anomalous, the anomaly can always be rotation into a single $U(1)_A$ while the remaining orthogonal $U(1)$'s all become non-anomalous. This anomalous $U(1)_A$ is broken by the
Green--Schwarz--Dine--Seiberg--Witten mechanism \cite{dsw}
in which a potentially large Fayet--Iliopoulos $D$--term
$\xi$ is generated by the VEV of the dilaton field.  (For the class of free fermionic models under investigation $ \xi > 0 $.) 
Such a $D$--term would, in general, break supersymmetry, unless
there is a direction $\hat\phi=\sum\alpha_i\phi_i$ in the scalar
potential for which $\sum Q_A^i\vert\alpha_i\vert^2$ is of opposite sign to
$\xi$ and that
is $D$--flat with respect to all the non--anomalous gauge symmetries,
as well as $F$--flat. 
If such a direction
exists, it will acquire a VEV, canceling the Fayet--Iliopoulos
$\xi$--term, restoring supersymmetry and stabilizing the vacuum.  

The $D$--term is formally defined as:
\begin{equation}
D_a^\alpha \equiv K_\alpha+ \sum_m \phi_m^\dagger T_a^\alpha \phi_m,
\end{equation}
where $T_a^\alpha$ is a generator and $\phi_m$ is a representation of the group, $\mathcal{G}_\alpha$  while the $K$--terms contain fields $\Phi_i$ like squarks, sleptons and Higgs bosons whose VEVs should vanish at this scale for good phenomenology.  When the $\phi_i$ are singlets of all the non--Abelian gauge groups in a model, the set of $D$- and $F$-flat constraints are then given by

\beqn
&& \langle D_A\rangle=\langle D_\alpha\rangle= 0~;\quad
\langle F_i\equiv
{{\partial W}\over{\partial\Phi_i}}\rangle=0~~;\label{dterms}\\
\nonumber\\
&& D_A=\left[K_A+
\sum Q_A^k\vert\phi_k\vert^2+\xi\right]~~;\label{da}\\
&& D_\alpha=\left[K_\alpha+
\sum Q_\alpha^k\vert\phi_k\vert^2\right]~,~\alpha\ne A~~;\label{dalpha}\\
&& \xi={{g^2({\rm Tr} Q_A)}\over{192\pi^2}}M_{\rm Pl}^2~~;
\label{dxi}
\eeqn
where $\chi_k$ are the fields which acquire VEVs of order
$\sqrt\xi$. The $\Phi_i$ is the superfield formed from the spacetime scalar $\phi_i$ and the chiral spin-1/2 fermion $\psi_m$, 
while $Q_A^k$ and $Q_\alpha^k$ denote the anomalous
and non--anomalous charges
and $M_{\rm Pl}\approx2\times 10^{18}$ GeV denotes the
reduced Planck mass. The solution ({\it i.e.}\  the choice of fields
with non--vanishing VEVs) to the set of
equations (\ref{dterms})--(\ref{dalpha}),
though nontrivial, is not unique. Therefore in a typical model there exist
a moduli space of solutions to the $F$- and $D$-flatness constraints,
which are supersymmetric and degenerate in energy \cite{moduli}. Much of
the study of the superstring model phenomenology 
(as well as non--string supersymmetric models \cite{savoy})
involves the analysis and classification of these flat directions.
The methods for this analysis in string models have been systematized
in \cite{systematic1,systematic2,cfn,mshsm}.

In general it is assumed in the literature that in a given string model
there should exist a supersymmetric solution to the $F$- and $D$-
flatness constraints. The simpler type of solutions utilize only
fields that are singlets of all the non--Abelian groups in a given
model (type I solutions). More involved solutions (type II)
that utilize non--Abelian fields have also been considered
\cite{mshsm}, as well as inclusion of non--Abelian fields
in systematic methods of analysis \cite{mshsm}.
The general expectation that a given model admits a supersymmetric 
solution arises from analysis of supersymmetric point quantum field theories.
In these cases it is known that if supersymmetry is preserved at the 
classical level, {\it i.e.} tree--level in perturbation theory, 
then there exist index theorems that forbid supersymmetry breaking 
at the perturbative quantum level \cite{witten1982}.
Therefore in point quantum field theories supersymmetry breaking
may only be induced by non--perturbative effects \cite{is}.

The issue of supersymmetry breaking in the string models 
that were constructed in refs. \cite{fmt,cfmt} merits further 
investigation. 
The aim of these models was to construct SM--like string
models with reduced untwisted Higgs spectrum. This was achieved 
by utilizing asymmetric boundary conditions in a basis vector
that does not break the $SO(10)$ symmetry \cite{fmt}. 
The consequence is that the 
entire vectorial {\bf 10} representation from the corresponding 
plane is projected out. The alternative is to utilize a combination 
of symmetric and asymmetric boundary conditions in basis vectors
that break the $SO(10)$ symmetry to the Pati--Salam gauge group
\cite{cfmt}. 
An unforeseen consequence of the Higgs reduction mechanism of 
refs.\  \cite{fmt,cfmt}
was the simultaneous projection of untwisted $SO(10)$ singlet fields. 
Consequently, the moduli space of supersymmetric flat
solutions is vastly reduced. In ref. \cite{fmt} it was concluded that the model under
investigation there does not 
contain supersymmetric flat directions that do not break some
of the SM symmetries.

For a given set of flat $D$-term VEVs, an $F$-term will break supersymmetry if its expectation value is non-zero. The scale of supersymmetry breaking is a function of the number of fields in the lowest-order, non-zero component of an $F$-term. Specifically, the lower the number of fields, the higher the energy scale of SUSY breaking. For example, a surviving second order component of an $F$-term breaks SUSY at the string/Planck scale, whereas a $17^{th}$ order component breaks as needed around the 100 GeV to 1 TeV scale.  This SUSY-breaking energy scale is decreased approximately 10 GeV per order increase of the non-zero $F$-term components.

In ref. \cite{cfmt} the issue of superymmetry breaking was
investigated further. It was shown that the model studied
in \cite{cfmt} does not have  $D$-flat directions that can
be proven to be $F$-flat to all order, 
other than through order-by-order analysis. 
That is, there do not appear to be any $D$-flat directions with {\it 
stringent} $F$-flatness (as defined in \cite{cfn,cfnw,Cleaver:2007ek}).
The analysis of the flat directions included all the fields in the
string model, {\it i.e.} SM singlet states as well as
SM charged states. The model of ref. \cite{cfmt}
therefore does not contain a $D$--flat direction that is also 
stringently $F$--flat to all order of non--renormalizable terms. 
The model may of course still admit non-stringent flat directions that rely 
on cancellations between superpotential terms. However, past experience
suggests that non--stringent flat directions
can only hold order by order, and are not maintained to all orders
\cite{pastexperience,cfs}.
This is a key difference between the string theory case, in which 
heavy string modes
generate an infinite tower of terms, versus the field theory case in which
heavy modes are not integrated out.
It was therefore speculated that in this case supersymmetry is not exact, 
but is in general broken at some order.

The SM--like string models presented in refs \cite{fmt,cfmt} 
contain three chiral generations,
charged under the SM gauge group and with the canonical
$SO(10)$ embedding of the weak--hypercharge, one pair of untwisted
electroweak Higgs doublets and a cubic level top--quark Yukawa coupling.
These string models therefore share some of the phenomenological
characteristics of the quasi--realistic free fermionic string models.
It may therefore represent an example of a quasi--realistic string model,
in which supersymmetry is broken due to the existence of the 
heavy string modes. The utilization of asymmetric boundary conditions
in these models entails that the geometrical moduli in these models are
projected out \cite{modulifixing}. 
The Higgs reduction mechanism of refs. \cite{fmt,cfmt}
further reduces the supersymmetric moduli space, and possibly
fixes it completely, while the hidden sector satifies the conditions for the dilaton race--track stabilization mechanism \cite{racetrack, cfmt}. It is noted that imposing more 
restrictive phenomenological constraints is correlated with
further reduction of the moduli space. The question of interest
is therefore whether the restriction of the supersymmetric moduli
space is specific to the case of standard--like models. In the next section
we investigate the Higgs reduction  mechanism in a model with an 
unbroken $SO(10)$ symmetry and in three flipped $SU(5)$ models. 

%
%
%
%

\section{The String Models}\label{stringmodelsection}


In this section we present the $SO(10)$ and flipped $SU(5)$ string
models that utilize the Higgs reduction mechanism of \cite{fmt,cfmt}.
We first review how the Higgs doublet--triplet splitting mechanism 
operates in the free fermionic models. For concreteness 
we illustrate how the mechanism operates in the model
of ref. \cite{fny}.

\subsection{Higgs Doublet--Triplet Splitting}\label{hdts}

An example of a free fermionic SM--like model is given in Table
\ref{fnymodel}\footnote{Only
the boundary condition basis vectors beyond the NAHE--set
are displayed.} .
\beqn
 &\begin{tabular}{c|c|ccc|c|ccc|c}
 ~ & $\psi^\mu$ & $\chi^{12}$ & $\chi^{34}$ & $\chi^{56}$ &
        $\bar{\psi}^{1,...,5} $ &
        $\bar{\eta}^1 $&
        $\bar{\eta}^2 $&
        $\bar{\eta}^3 $&
        $\bar{\phi}^{1,...,8} $ \\
\hline
\hline
  ${b_4}$     &  1 & 1&0&0 & 1~1~1~1~1 & 1 & 0 & 0 & 0~0~0~0~0~0~0~0 \\
  ${\alpha}$   &  1 & 0&0&1 & 1~1~1~0~0 & 1 & 1 & 0 & 1~1~1~1~0~0~0~0 \\
  ${\beta}$  &  1 & 0&1&0 &
                ${1\over2}$~${1\over2}$~${1\over2}$~${1\over2}$~${1\over2}$
              & ${1\over2}$ & ${1\over2}$ & ${1\over2}$ &
                1~1~${1\over2}$~0~${1\over2}$~${1\over2}$~${1\over2}$~0 \\
\end{tabular}
   \nonumber\\
   ~  &  ~ \nonumber\\
   ~  &  ~ \nonumber\\
     &\begin{tabular}{c|c|c|c}
 ~&   $y^3{y}^6$
      $y^4{\bar y}^4$
      $y^5{\bar y}^5$
      ${\bar y}^3{\bar y}^6$
  &   $y^1{\omega}^6$
      $y^2{\bar y}^2$
      $\omega^5{\bar\omega}^5$
      ${\bar y}^1{\bar\omega}^6$
  &   $\omega^1{\omega}^3$
      $\omega^2{\bar\omega}^2$
      $\omega^4{\bar\omega}^4$
      ${\bar\omega}^1{\bar\omega}^3$ \\
\hline
\hline
$b_4$ & 1 ~~~ 0 ~~~ 0 ~~~ 1  & 0 ~~~ 0 ~~~ 1 ~~~ 0  & 0 ~~~ 0 ~~~ 1 ~~~ 0 \\
$\alpha$  & 0 ~~~ 0 ~~~ 0 ~~~ 1  & 0 ~~~ 1 ~~~ 0 ~~~ 1  & 1 ~~~ 0 ~~~ 1 ~~~ 0
\\
$\beta$ & 0 ~~~ 0 ~~~ 1 ~~~ 1  & 1 ~~~ 0 ~~~ 0 ~~~ 1  & 0 ~~~ 1 ~~~ 0 ~~~ 0 \\
\end{tabular}
\label{fnymodel}
\eeqn

The Higgs doublet--triplet splitting operates as follows \cite{ps}.
The Neveu--Schwarz sector gives rise to three fields in the {\bf 10} representation of $SO(10)$.  These contain the  Higgs electroweak
doublets and color triplets. Each of those is charged with respect to one
of the horizontal $U(1)$ symmetries $U(1)_{1,2,3}$.  Each one of these
multiplets is associated, by the horizontal symmetries, with one of the
twisted sectors, $b_1$, $b_2$ and $b_3$. The doublet--triplet
splitting results from the boundary condition basis vectors which break
the $SO(10)$ symmetry to $SO(6)\times SO(4)$. We can define a quantity
$\Delta_i$ in these basis vectors which measures the difference between the
boundary conditions assigned to the internal fermions from the set
$\{y,w\vert{\bar y},{\bar\omega}\}$ and which are periodic in the vector
$b_i$,
\begin{equation}
\Delta_i=\vert\alpha_L({\rm internal})-
\alpha_R({\rm internal})\vert=0,1~~(i=1,2,3)
\label{dts}.
\end{equation}
If $\Delta_i=0$ then the Higgs triplets, $D_i$ and ${\bar D}_i$,
remain in the massless spectrum while the Higgs doublets, $h_i$ and
${\bar h}_i$, are projected out
and the opposite occurs for $\Delta_i=1$.

The rule in Eq. (\ref{dts}) is a generic rule that operates in
free fermionic models.
The model of eq. (\ref{fnymodel}) illustrates this rule.
In this model the basis vector that breaks $SO(10)$ symmetry to $SO(6)\times
SO(4)$ is $\alpha$ and, with respect to $\alpha$,
$\Delta_1=\Delta_2=\Delta_3=1$. Therefore,
this model produces three pairs of electroweak
Higgs doublets from the Neveu--Schwarz sector, $h_1$, $\bar h_1$
$h_2$, $\bar h_2$ and $h_3$, $\bar h_3$, and all the untwisted color
triplets are projected out. Note also that the vector basis $b_4$ is symmetric
with respect to the internal fermions
that are periodic in the vectors $b_i, \ i=1,2,3$ and, therefore, does not
project out the fields in the {\bf 10} representation of $SO(10)$.

Another possibility is to construct models in which
both the Higgs color triplets and electroweak doublets from the
Neveu--Schwarz sector are projected out by the GSO projections.
This is a viable possibility as we can choose for example
$$\Delta_j^{(\alpha)}=1 ~{\rm and}~ \Delta_j^{(\beta)}=0,$$
where $\Delta^{(\alpha,\beta)}$ are the projections due
to the basis vectors $\alpha$ and $\beta$ respectively.
This is desirable as the number of Higgs representations,
which generically appear in the massless spectrum,
is larger than what is allowed by the low energy phenomenology.
The model shown in eq. (\ref{stringmodel})
illustrates this possibility

\beqn
 &\begin{tabular}{c|c|ccc|c|ccc|c}
 ~ & $\psi^\mu$ & $\chi^{12}$ & $\chi^{34}$ & $\chi^{56}$ &
        $\bar{\psi}^{1,...,5} $ &
        $\bar{\eta}^1 $&
        $\bar{\eta}^2 $&
        $\bar{\eta}^3 $&
        $\bar{\phi}^{1,...,8} $ \\
\hline
\hline
  ${\alpha}$  &  0 & 0&0&0 & 1~1~1~0~0 & 1 & 0 & 0 & 1~1~0~0~0~0~0~0 \\
  ${\beta}$   &  0 & 0&0&0 & 1~1~1~0~0 & 0 & 1 & 0 & 0~0~1~1~0~0~0~0 \\
  ${\gamma}$  &  0 & 0&0&0 &
		${1\over2}$~${1\over2}$~${1\over2}$~${1\over2}$~${1\over2}$
	      & ${1\over2}$ & ${1\over2}$ & ${1\over2}$ &
                0~0~0~0~$1\over2$~$1\over2$~${1\over2}$~${1\over2}$ \\
\end{tabular}
   \nonumber\\
   ~  &  ~ \nonumber\\
   ~  &  ~ \nonumber\\
     &\begin{tabular}{c|c|c|c}
 ~&   $y^3{y}^6$
      $y^4{\bar y}^4$
      $y^5{\bar y}^5$
      ${\bar y}^3{\bar y}^6$
  &   $y^1{\omega}^5$
      $y^2{\bar y}^2$
      $\omega^6{\bar\omega}^6$
      ${\bar y}^1{\bar\omega}^5$
  &   $\omega^2{\omega}^4$
      $\omega^1{\bar\omega}^1$
      $\omega^3{\bar\omega}^3$
      ${\bar\omega}^2{\bar\omega}^4$ \\
\hline
\hline
$\alpha$ & 1 ~~~ 0 ~~~ 0 ~~~ 1  & 0 ~~~ 0 ~~~ 1 ~~~ 1  & 0 ~~~ 0 ~~~ 1 ~~~ 1 \\
$\beta$  & 0 ~~~ 0 ~~~ 1 ~~~ 1  & 1 ~~~ 0 ~~~ 0 ~~~ 1  & 0 ~~~ 1 ~~~ 0 ~~~ 1 \\
$\gamma$ & 0 ~~~ 1 ~~~ 0 ~~~ 0  & 0 ~~~ 1 ~~~ 0 ~~~ 0  & 1 ~~~ 0 ~~~ 0 ~~~ 0 \\
\end{tabular}
\label{stringmodel}
\eeqn

Both the basis vectors $\alpha$ and $\beta$ break the 
$SO(10)$ symmetry to $SO(6)\times SO(4)$ and the basis vector
$\gamma$ breaks it further to $SU(3)\times U(1)_C\times SU(2)\times U(1)_L$.  
The basis vector $\alpha$ is symmetric with respect to the sector
$b_1$ and asymmetric with respect to the sectors $b_2$ and $b_3$, 
whereas the basis vector $\beta$ is symmetric with respect to $b_2$ 
and asymmetric with respect to $b_1$ and $b_3$. As a consequence of these
assignments and of the
string doublet--triplet splitting mechanism discussed above,
both the untwisted Higgs color triplets and electroweak doublets,
with leading coupling to the matter states from the sectors 
$b_1$ and $b_2$, are projected out by the generalized GSO projections. 
At the same time the untwisted color Higgs triplets 
that couple at leading order to the states from the sector $b_3$ are projected out, 
whereas the untwisted electroweak Higgs doublets
remain in the massless spectrum. Due to the asymmetric boundary 
conditions in the sector $\gamma$ with respect to the sector
$b_3$, the leading Yukawa coupling is that of the up--type quark from the
sector $b_3$ to the untwisted electroweak Higgs doublet \cite{top}.
Hence, the leading Yukawa term is that of the top quark and only its
mass is characterized by the electroweak VEV \cite{top}. The lighter
quarks and leptons couple to the light Higgs doublet through
higher order non-renormalizable operators that become effective 
renormalizable operators by the VEVs that are used to cancel the
anomalous $U(1)_A$ $D$--term equation \cite{top}. The novelty
in the construction of (\ref{stringmodel}) 
is that the reduction of the untwisted
Higgs spectrum is obtained by the choice of the boundary
condition basis vectors in eq. (\ref{stringmodel}), 
without resorting to analysis of supersymmetric flat directions,
whereas in other models it is obtained by the choice of
flat directions and analysis of the superpotential \cite{reviewffm}.

However, the surprising result was that the model does not seem to admit 
a stringent supersymmetric solution \cite{cfmt}. This appeared to be a consequence of the reduction of untwisted singlet states (denoted by $\Phi_i$ and $\phi_i$ in that model and in all models presented here), simultaneous with the untwisted Higgs reduction, imposed by the asymmetric/symmetric boundary conditions.  We next turn to examine 
this issue in models in which the $SO(10)$ symmetry remains unbroken, 
or is broken to the flipped $SU(5)$ subgroup. 

\subsection{An $SO(10)$ Model}\label{so10model}
In the previous section we discussed the Higgs reduction mechanism 
of ref. \cite{fmt,cfmt}. We noted that the reduction is achieved 
by utilizing symmetric and asymmetric boundary conditions in two 
separate basis vectors. However, the reduction can be obtained
if we assign asymmetric boundary conditions in a basis vector 
that does not break the $SO(10)$ symmetry. The model in 
Table \ref{so10model} is an example of such a model.  


\begin{table}[h!t!p!]
\begin{center}
 \begin{tabular}{c|c|ccc|c|ccc|c}
 ~ & $\psi^\mu$ & $\chi^{12}$ & $\chi^{34}$ & $\chi^{56}$ &
        $\bar{\psi}^{1,...,5} $ &
        $\bar{\eta}^1 $&
        $\bar{\eta}^2 $&
        $\bar{\eta}^3 $&
        $\bar{\phi}^{1,...,8} $ \\
\hline
\hline
  ${b_4   }$  &  0 & 0&0&0 & 1~1~1~1~1 & 1 & 0 & 0 & 0~0~0~0~0~0~0~0 \\
  ${b_5   }$  &  0 & 0&0&0 & 1~1~1~1~1 & 0 & 1 & 0 & 0~0~0~0~0~0~0~0 \\
  ${b_6   }$  &  0 & 0&0&0 & 1~1~1~1~1 & 1 & 1 & 0 & 1~1~0~0~0~0~0~0  \\
  ${2\gamma}$  &  0 & 0&0&0 & 1~1~1~1~1 & 1 & 1 & 1 & 1~0~1~1~1~0~0~0  \\
\end{tabular}

\vspace{20pt}

\begin{tabular}{c|c|c|c|c|c|c}
 ~&   $y^{3 \dots 6}$
    &   ${\bar y}^{3 \dots 6}$
    &   $y^{1,2} \   \omega^{5,6}$   
    &   ${\bar y}^{1,2} \   {\bar \omega}^{5,6}$   
  &   $\omega^{1 \ldots 4}$ 
  &  ${\bar\omega}^{1 \ldots 4}$ \\
\hline
\hline
${b_4}   $  & 1~0~0~1&1~0~0~1&0~0~0~1&1~0~1~1&0~0~1~0&0~1~1~1\\
${b_5}  $   & 0~0~1~0&1~0~1~1&1~0~1~0&1~0~1~0&1~0~0~0&1~1~0~1\\
${b_6} $  & 0~1~0~0&0~1~0~0&0~1~0~0&0~1~0~0&0~1~0~1&0~0~0~0\\
$2\gamma$ & 0~0~0~0&0~0~0~0&0~0~0~0&0~0~0~0&0~0~0~0&0~0~0~0\\
\end{tabular}
\end{center}
\caption{Additional boundary vectors for the $SO(10)$ model}
\label{so10model}
\end{table}

\begin{table}
\begin{equation*}
{\bordermatrix{
         &{\bf 1}&  S & &{b_1}&{b_2}&{b_3}& &{b_4}&{b_5}&{2\gamma}& \delta\cr
 {\bf 1} &   0 &0 & & 0  &  1 & 1  & &  0&  0 &  0&  0     \cr
       S &   0 &0 & &0  & 0 &0  & &  0&  0 &  0&  0     \cr
         &       &    & &     &     &     & &    &     &    &         \cr
   {b_1} &    0 & 1 & & 0  &  1 & 1  & &  0&  0 & 1& 1     \cr
   {b_2} &    1 & 1 & & 1  &  1 & 1  & &  0& 1 & 1& 1     \cr
   {b_3} &    1 & 1 & & 1  &  1 & 1  & & 1&  0 & 1& 1     \cr
	 &       &    & &     &     &     & &    &     &    &         \cr
{b_4} &    0 & 0 & &1  & 1 & 0  & & 0&  1 & 1& 0     \cr
{b_5} &    0 & 0 & &1  &  0 &1  & &  1& 0 & 0& 1     \cr
{ 2\gamma  } &    0 & 0 & & 0  &  0 &1  & &  0&  1 & 0& 1     \cr
{\delta}&    0 & 0 & & 0  &  0 & 0  & &  1&  0 & 1&  1     \cr}}
\end{equation*} 
\caption{GSO K--Matrix for the $SO(10)$ model.  The K$_{ij}$ components are labeled by column `j' and row `i'.  All other GSO K--matrix tables follow the same pattern.}
\label{phasesomodel}
\end{table}

We note that in tables \ref{so10model} and \ref{phasesomodel} vectors that do not
break the $SO(10)$ symmetry are denoted by $b_j$ with $j\ge4$. Basis vectors that 
break the $SO(10)$ symmetry in the fliipped $SU(5)$ models below will be
denoted by Greek letters. The last vector in tables \ref{so10model} and \ref{phasesomodel}
is denoted as $2\gamma$ to adhere with the notation used in the quasi--realistic 
free fermionic models \cite{foc}.

The model defined by Tables \ref{so10model} and \ref{phasesomodel}
contain three spinorial {\bf 16} representations of the $SO(10)$ GUT group.
All untwisted vectorial {\bf 10} representations are projected out. The asymmetric
boundary conditions in $b_4$ with respect to $b_2$ and $b_3$ projects out
the corresponding vectorial representations, whereas the asymmetric boundary
condition in $b_5$ with respect to $b_1$ project out the remaining untwisted
$SO(10)$ vectorial representations. The twisted sectors $b_1+b_4$ and $b_1+b_2+b_4+b_5$
give rise to $SO(10)$ vectorial representations and these are the sources of the two pairs of SM Higgs: $h_1 ({\bar h}_1)$ and $h_2 ({\bar h}_2)$. 

We note that the model in Tables \ref{so10model} and \ref{phasesomodel} is not phenomenologically realistic 
as it does not contain the heavy Higgs representations needed to break the
$SO(10)$ symmetry to the SM gauge group. Our interest in this
model here is to study the existence and characteristic of the (all-order) stringent 
flat directions. Namely, do they exist, and do they preserve the 
SM (within the $SO(10)$) gauge group?  This question
will be discussed in section \ref{flatdir}.  The states and gauge charges for this model can be found in Table \ref{so10gauge} located in Appendix A.
\newpage
\subsection{Flipped $SU(5)$ Models}
Some potentially more realistic models are those with flipped $SU(5)$ observable sectors.  This are listed below in sections: \ref{su51model1}, \ref{su51model2} and \ref{su51model3}.

\subsubsection{Flipped $SU(5)$ Model 1}\label{su51model1}


This flipped $SU(5)$ model has additional boundary vectors given in Table \ref{fsu5model1} and a GSO projection matrix shown in Table \ref{phasesu51model1}. The states and gauge charges for this model can be found in Table \ref{su51gauge1} located in Appendix A. This model is completely lacking in both untwisted and twisted Higgs.

\begin{table}[!h]
\begin{center}
 \begin{tabular}{c|c|ccc|c|ccc|c}
 ~ & $\psi^\mu$ & $\chi^{12}$ & $\chi^{34}$ & $\chi^{56}$ &
        $\bar{\psi}^{1,...,5} $ &
        $\bar{\eta}^1 $&
        $\bar{\eta}^2 $&
        $\bar{\eta}^3 $&
        $\bar{\phi}^{1,...,8} $ \\
\hline
\hline
  ${b_4}$  &  0 & 0&0&0 & 1~1~1~1~1 & 1 & 0 & 0 & 0~0~0~0~0~0~0~0 \\
  ${b_5}$   &  0 & 0&0&0 & 1~1~1~1~1 & 1 & 0 & 0 & 0~0~0~0~0~0~0~0 \\
  ${\gamma}$  &  0 &0&0&0 &${1\over 2}$~${1\over 2}$~${1\over 2}$~${1\over 2}$~${1\over 2}$& ${1\over 2}$ & ${1\over 2}$ & ${1\over 2}$ & 0~0~0~0~${1\over 2}$~${1\over 2}$~${1\over 2}$~${1\over 2}$ \\
\end{tabular}
\vspace{20pt}

\begin{tabular}{c|c|c|c|c|c|c}
 ~&   $y^{3 \dots 6}$
    &   ${\bar y}^{3 \dots 6}$
    &   $y^{1,2} \   \omega^{5,6}$   
    &   ${\bar y}^{1,2} \   {\bar \omega}^{5,6}$   
  &   $\omega^{1 \ldots 4}$ 
  &  ${\bar\omega}^{1 \ldots 4}$ \\
\hline
\hline
${b_4} $  & 1~0~0~1&1~0~0~1&0~0~0~1&1~0~1~1&0~0~1~0&0~1~1~1\\
${b_5} $   & 0~0~1~0&1~0~1~1&1~0~1~0&1~0~1~0&1~0~0~0&1~1~0~1\\
${\gamma}$& 0~1~0~0&0~1~0~0&0~1~0~0&0~1~0~0&0~1~0~1&0~0~0~0\\
\end{tabular}
\end{center}
\caption{Additional boundary vectors for flipped $SU(5)$ model \#1}
\label{fsu5model1}
\end{table}

\begin{table}[!h]
\begin{center}
\begin{equation*}
{\bordermatrix{
        &{\bf 1}&  S & &{b_1}&{b_2}&{b_3}& &{b_4}&{b_5}&~~{\gamma}\cr
 {\bf 1}&   0 &0 & & 1  &  1 & 1  & &  0    &  0   & -{1\over2}   \cr
       S&   0 &0 & &0  & 0 &0 & & 0    & 0   & ~~ 0   \cr
        &       &    & &     &     &     & &        &       &       \cr
   {b_1}&   1 & 1 & & 1  &  1 & 1  & & 0    &0   & -{1\over 2 }  \cr
   {b_2}&   1 & 1 & & 1  &  1 & 1  & & 0    &1   & ~~{1\over 2}  \cr
   {b_3}&   1 & 1 & & 1  &  1 & 1  & & 1    &0   & -{1\over 2 }  \cr
	&       &    & &     &     &     & &        &       &       \cr
{b_4}&    0 & 0 & & 1  &  1 & 0  & &  0    &1   & -{1\over 2}   \cr
 {b_5}&    0 & 0 & & 1  &  0 & 1  & &  1    &0   & ~~{1\over 2}   \cr
{\gamma}&  0 & 0 & & 1  &  0 & 0  & &  0    &0   & -{1\over 2}   \cr}}
\end{equation*}
\end{center}
\caption{GSO K-Matix for Flipped $SU(5)$ Model \#1}
\label{phasesu51model1}
\end{table}
\newpage
\subsubsection{Flipped $SU(5)$ model 2}\label{su51model2}


This flipped $SU(5)$ model has additional boundary vectors given in Table \ref{fsu5model2} and a GSO projection matrix shown in Table \ref{phasesu51model2}. The states and gauge charges for this model can be found in Table \ref{su51gauge2} located in Appendix A. This model has two untwisted Higgs pairs and three more pairs of twisted Higgs.

\begin{table}[!h]
\begin{center}
 \begin{tabular}{c|c|ccc|c|ccc|c}
 ~ & $\psi^\mu$ & $\chi^{12}$ & $\chi^{34}$ & $\chi^{56}$ &
        $\bar{\psi}^{1,...,5} $ &
        $\bar{\eta}^1 $&
        $\bar{\eta}^2 $&
        $\bar{\eta}^3 $&
        $\bar{\phi}^{1,...,8} $ \\
\hline
\hline
  ${b_4}$  &  1 & 0&0&0 & 1~1~1~1~1 & 0 & 1 & 0 & 0~0~0~0~0~0~0~0 \\
  ${b_5}$   &  1 & 0&1&0 & 1~1~1~1~1 & 1 & 0 & 0 & 0~0~0~0~0~0~0~0 \\
  ${\gamma}$  &  1 & 0&0&1 & ${1\over 2}$~${1\over 2}$~${1\over 2}$~${1\over 2}$~${1\over 2}$ & ${1\over 2}$& ${1\over 2}$&${1\over 2}$&${1\over 2}$~${1\over 2}$~${1\over 2}$~${1\over 2}$~${1\over 2}$~${1\over 2}$~${1\over 2}$~${1\over 2}$ \\
\end{tabular}

\vspace{20pt}

\begin{tabular}{c|c|c|c|c|c|c}
 ~&   $y^{3 \dots 6}$
    &   ${\bar y}^{3 \dots 6}$
    &   $y^{1,2} \   \omega^{5,6}$   
    &   ${\bar y}^{1,2} \   {\bar \omega}^{5,6}$   
  &   $\omega^{1 \ldots 4}$ 
  &  ${\bar\omega}^{1 \ldots 4}$ \\
\hline
\hline
${b_4}   $  & 1~0~0~1&0~0~0~0&0~0~1~0&1~0~1~1&0~0~0~1&0~0~0~1\\
${b_5} $   & 0~0~0~0&1~0~0~1&1~0~1~1&0~0~1~0&0~1~0~0&0~1~0~0\\
${\gamma}  $  & 0~1~0~0&1~1~0~1&0~1~0~0&0~1~0~0&1~0~1~0&0~0~0~0\\
\end{tabular}
\end{center}
\caption{Additional boundary vectors for $SU(5)$ model \#2}
\label{fsu5model2}
\end{table}

\begin{table}[!h]
\begin{center}
\begin{equation*}
{\bordermatrix{
        &{\bf 1}&  S & &{b_1}&{b_2}&{b_3}& &{b_4}&{b_5}&~~{\gamma}\cr
 {\bf 1}&  0 &0& & 1  &  1 & 1  & &  1    &  1   & ~~1\cr
       S&   0&0& &0& 0&0& &0& 0& ~~ 0\cr
        &       &    & &     &     &     & &        &       &       \cr
   {b_1}&    1 & 1 & & 1  &  1 & 1  & & 0& 0&~~ {1\over 2}\cr
   {b_2}&    1 & 1 & & 1  &  1 & 1  & & 0& 0& ~~1\cr
   {b_3}&    1 & 1 & & 1  &  1 & 1  & &  0& 1 &~~ 0\cr
	&       &    & &     &     &     & &        &       &       \cr
{b_4}&    1 & 1 & & 1  &  1 & 0 & &  1    & 1   & ~~1\cr
{b_5}&    1 & 1 & & 1  &  1 & 1  & & 1    &  1   & -{1\over 2}\cr
{\gamma}&   0&1 & &0& 0& 0& & 0& 1&~~ {1\over 2}\cr}}
\end{equation*}
\end{center}
\caption{GSO K-Matix for Flipped $SU(5)$ Model \#2}
\label{phasesu51model2}
\end{table}

\newpage
\subsubsection{Flipped $SU(5)$ Model 3}\label{su51model3}


This flipped $SU(5)$ model has additional boundary vectors given in Table \ref{fsu5model3} and a GSO projection matrix shown in Table \ref{phasesu51model3}. The states and gauge charges for this model can be found in Table \ref{su51gauge3} located in Appendix A.
\begin{table}[!h]
\begin{center}
 \begin{tabular}{c|c|ccc|c|ccc|c}
 ~ & $\psi^\mu$ & $\chi^{12}$ & $\chi^{34}$ & $\chi^{56}$ &
        $\bar{\psi}^{1,...,5} $ &
        $\bar{\eta}^1 $&
        $\bar{\eta}^2 $&
        $\bar{\eta}^3 $&
        $\bar{\phi}^{1,...,8} $ \\
\hline
\hline
  ${b_4}$  &  1 & 1&0&0 & 1~1~1~1~1 & 0 & 1 & 0 & 0~0~0~0~0~0~0~0 \\
  ${b_5}$   &  1 & 0&1&0 & 1~1~1~1~1 & 1 & 0 & 0 & 0~0~0~0~0~0~0~0 \\
  ${\gamma}$  &  1& 0&0&1 &${1\over 2}$~${1\over 2}$~${1\over 2}$~${1\over 2}$~${1\over 2}$& ${1\over 2}$&${1\over 2}$& ${1\over 2}$&1~1~${1\over 2}$~${1\over 2}$~${1\over 2}$~${1\over 2}$~0~0 \\
\end{tabular}

\vspace{20pt}

\begin{tabular}{c|c|c|c|c|c|c}
 ~&   $y^{3 \dots 6}$
    &   ${\bar y}^{3 \dots 6}$
    &   $y^{1,2} \   \omega^{5,6}$   
    &   ${\bar y}^{1,2} \   {\bar \omega}^{5,6}$   
  &   $\omega^{1 \ldots 4}$ 
  &  ${\bar\omega}^{1 \ldots 4}$ \\
\hline
\hline
${b_4}   $  & 1~0~0~1&0~0~0~0&0~0~1~0&1~0~1~1&0~0~0~1&0~0~0~1\\
${b_5} $   & 0~0~0~0&1~0~0~1&1~0~1~1&0~0~1~0&0~1~0~0&0~1~0~0\\
${\gamma}  $  & 0~1~0~0&1~1~0~1&0~0~0~0&1~0~0~1&1~1~1~0&0~1~0~0\\
\end{tabular}
\end{center}
\caption{Additional boundary vectors for $SU(5)$ model \#3}
\label{fsu5model3}
\end{table}

\begin{table}[!h]
\begin{center}
\begin{equation*}
{\bordermatrix{
        &{\bf 1}&  S & &{b_1}&{b_2}&{b_3}& &{b_4}&{b_5}&~~{\gamma}\cr
 {\bf 1}&   0&0& &1  &  1 & 1  & & 0& 0& -{1\over 2}   \cr
       S&   0 &0& &0& 0&0& & 1& 1   &  ~~0   \cr
        &       &    & &     &     &     & &        &       &       \cr
   {b_1}&    1 & 1 & & 1  &  1 & 1  & & 1    & 1   & -{1\over 2}    \cr
   {b_2}&    1 & 1 & & 1  &  1 & 1  & & 1    & 1   & ~~{1\over 2}    \cr
   {b_3}&    1 & 1 & & 1  &  1 & 1  & & 1    & 1   & ~~0 \cr
	&       &    & &     &     &     & &        &       &       \cr
{b_4}&    0 & 0& &0&  0& 1& &  0& 1   & ~~{1\over 2}   \cr
 {b_5}&    0 & 0& &0&  0& 1& &  1& 0   & ~~{1\over 2}   \cr
{\gamma}&   1 &1 & &1  & 1 & 0& & 1    & 1   & ~~1   \cr}}
\end{equation*}
\end{center}
\caption{GSO K-Matix for Flipped $SU(5)$ Model \#3}
\label{phasesu51model3}
\end{table}

This last model produces three generations from the sectors $b_1$, $b_2$ and $b_3$ and
one light Higgs pair from the untwisted sector. The sectors $b_4$ and $b_5$ give rise to 
heavy Higgs states that are needed to break the $SU(5)$ gauge symmetry to the 
Standard Model gauge group. The model admits a cubic level top quark Yukawa coupling of order
one. In this respect the model reproduces many of the features of other quasi--realistic 
flipped $SU(5)$ string models. Its distinctive characteristic is that two of 
the untwisted Higgs pairs are projected out due to the asymmetric boundary conditions
in $b_4$ and $b_5$ with respect to $b_1$ and $b_2$.

\section{Flat directions}\label{flatdir}
Here we list the results of the flat direction analysis.  These lists are by no means exhaustive but are presented just as a ``proof of concept''.  The $SO(10)$ model has $D$--flat directions found in Table \ref{so10dflat} and all-order stringent $F$--flat directions in Table \ref{so10fflat}.  For this $SO(10)$ model it was necessary only to investigate type I flat directions (those formed with fields that are singlets under both the observable and hidden sector gauge groups).  This is, of course, the preferred option as it is both simpler and computationally less-expensive. The all-order stringent flat directions we found possessed only VEVs of singlets fields and  since the SM model is completely contained within $SO(10)$ that means that  the SM remains unbroken here.  

Similarly, those flat directions investigated in flipped $SU(5)$ model \#2, were also of type I. The $D$-flat directions for this model can be found in Table \ref{su52dflat} and the $F$-flat table can be found in Table \ref{su52fflat}. We note that the all-order stringent flat directions in Table \ref{su52fflat} contain VEVs for one or more generations of anti-electrons.  Thus the SM hypercharge is broken in this model.  Interestingly, this model contains an exotic negatively-charged electron singlet forming a vector-pair with the corresponding anti-electron (see Table \ref{su51gauge2}).  

Both flipped $SU(5)$ models \#1 and \#3, however, lacked type I stringent flat directions and required an investigation of those fields which are charged under some non-Abelian gauge representation.  For model \#1, the failure to produce singlet $D$-flat directions that can solely cancel the Fayet-Iliopoulos term came as a result of all possible directions having an anomalous charge of zero (see Table \ref{su51dflatsing}). This condition requires the search for type II directions with negative anomalous charge.  This search is a nonlinear process (\cite{cfmt},\cite{Buchmuller:2006ik}) and, for the purposes of our investigation, the use of stringency requirements offered a much quicker, and still fruitful, search.  The net effect of stringent flatness is that: (1.) at least two fields (including a given field appearing twice) must not take on a VEV or (2.) self-cancellation between components of a non-abelian field must occur.  Additionally for model \#1, we found (as can be seen in Table \ref{su51dflatNA}) non-Abelian $D$-flat directions that were negative. However, these directions failed to yield $F$-flat directions by themselves; this necessitated the mixing of both singlet and non-Abelian $D$-flat directions in order to find all-order stringent $F$-flatness.  These mixed $D$-flat directions can be seen in Table \ref{su51dflatmix}.

The speed increase from stringent $F$-flat direction searches results from only needing to investigate a finite set of potentially dangerous superpotential terms which determine all-order flatness.  The answer to whether or not these terms are dangerous (i.e. they actually exist) comes from the fact that these models are derived from string theory and, therefore, the gauge invariant terms need to also obey Ramon/Neveu-Schwarz (RNS) worldsheet charge conservation.  This then makes string-derived models more constrained than the usual QFT-derived ones.  These all-order RNS rules can be found in \cite{mshsm}.

The $D$- and $F$-flat tables for flipped $SU(5)$ model \#1 can be found in Tables \ref{su51dflatsing}--\ref{su51fflat}. Here the $SU(5)$ gauge group remains unbroken.  However, hypercharge is most likely broken in this model since in flipped $SU(5)$ there is a contribution to hypercharge from an external $U(1)$ and all extra $U(1)'_i$ appear to be carried by the singlet and/or non-Abelian states acquiring VEVs in the all-order stringent F-flat directions (see Tables \ref{su51fflat} and \ref{su51gauge1}).  Additionally the hidden sector $SU(4)$ is broken by the ${\bar H}_i$ while the $SO(10)_{hidden}$ remains unbroken.  

For flipped $SU(5)$ model \#3, the $D$- and $F$-flat tables can be found in Tables  \ref{su53dflatsing}--\ref{su53fflat}. We find here that $SU(5)$ remains unbroken and it is unclear whether hypercharge is broken in the few all-order stringent flat directions calculated.  As for the hidden sector, we see from Table \ref{su51gauge3} that the first and third $SU(2)$'s are broken while the remaining $SU(2)^2 \times SU(4)$ is left unbroken.  

\small\addtolength{\tabcolsep}{-3pt}
\begin{table}[H]
\begin{center}
\begin{tabular}{l|r|rrrrrrrrrrrrrrrrr}
\hline \rule{0pt}{2.6ex}
&$Q_A$& $\phi_1$& $\phi_2$& $\phi_3$& $\psi_1$& $\psi_2$& $\psi_3$& $\psi_4$& $\psi_5$& $\psi_6$& 
$\psi_7$& $\psi_8$& $\psi_9$& $\psi_{10}$& $\psi_{11}$& $\psi_{12}$& $\psi_{13}$& $\psi_{14}$\\
\hline
${\cal D}_1$&   -1&  0&  0&  0&  0&  0&  0&  0&  0& -3&  0&  0&  0&  1&  2&  2& -2&  1\\
${\cal D}_2$&   -1&  0& -3&  0&  0&  0&  0&  0&  0&  0&  0&  0&  0& -2&  2&  2& -2&  4\\
${\cal D}_3$&   -1&  0&  0& -3&  0&  0&  0&  0&  0&  0&  0&  0&  0&  4&  2&  2& -2& -2\\
${\cal D}_4$&    0&  1&  0&  0&  0&  0&  0&  0&  0&  0&  0&  0&  0& -2&  0&  0&  0&  2\\
${\cal D}_5$&    0&  0&  0&  0&  0&  0&  0&  0&  0&  0&  1&  0&  0&  0&  0&  0& -1&  1\\
${\cal D}_6$&    0&  0&  0&  0&  0&  0&  0&  0&  0&  0&  0&  1&  0&  0&  1&  0&  0& -1\\
${\cal D}_7$&    0&  0&  0&  0&  1&  0&  0&  0&  0&  0&  0&  0&  0&  1&  0&  0& -1&  0\\
${\cal D}_8$&    0&  0&  0&  0&  0&  0&  0&  0&  0&  0&  0&  0&  1&  0&  0&  1&  0&  1\\
${\cal D}_9$&    0&  0&  0&  0&  0&  0&  0&  1&  0&  0&  0&  0&  0&  1&  0&  0&  0& -1\\
${\cal D}_{10}$& 0&  0&  0&  0&  0&  0&  0&  0&  1&  0&  0&  0&  0&  1&  0&  0&  0& -1\\
${\cal D}_{11}$& 0&  0&  0&  0&  0&  1&  0&  0&  0&  0&  0&  0&  0& -1&  1&  0&  0&  0\\
${\cal D}_{12}$& 0&  0&  0&  0&  0&  0&  1&  0&  0&  0&  0&  0&  0&  1&  0&  1&  0&  0\\
\hline
\end{tabular}\\
\end{center}
\caption{Basis set of U(1) singlet $D$-flat directions for the $SO(10)$ model. These directions were sufficient to produce all-order stringent $F$-flat directions. Column 1 denotes the $D$-flat direction label. Column 2 is the anomalous charge. The remaining columns specify the norm squared VEVs of the respective non-Abelian singlet fields.  All other $D$-flat tables are structured in a like manner.}
\label{so10dflat}
\end{table}

\begin{table}[H]
\begin{center}
\begin{tabular}{l|r|rrrrrrrrrrrr}
\hline \rule{0pt}{2.6ex}
FD &$Q_A$& $\Phi_1$& $\Phi_2$& $\Phi_3$& $\phi_3$& $\phi_6$& $\phi_8$& $\phi_9$& $\phi_{10}$& $\phi_{11}$& 
$\phi_{12}$& $\phi_{13}$& $\phi_{14}$\\
\hline
1&  -49&  0&  -3&  -144& -98&  0&  0&  0& 92&  98&  0&  -98&  -92\\
2&  -1&  0&  -1&  -2&  -2&  0&  0&  0&  0&  2&  0& -2&  0\\
3&  -1&  0&  0&  -3&  -4&  0&  0&  0&  0&  2&  -2&  -2& -2\\
4&  -1&  0&  -1&  0&  0&  -2&  0&  -2&  0&  2&  0&  -2& 0\\
5&  -2&  0&  -3&  -1&  0&  -2&  0&  -4&  0&  4&  0&  -4&  0\\
6&  -1&  0&  0&  1&  0&  -4&  0&  -2&  0&  2&  0&  -2&  0\\
7&  -1&  0&  0&  -1&  -2&  -2&  0&  0&  0&  2&  0&  -2&  0\\
8&  -2&  0&  -1&  -3&  -4&  -2&  0&  0&  0&  4&  0&  -4&  0\\
9&  -1&  0&  1&  0&  -2&  -4&  0&  0&  0&  2&  0&  -2&  0\\
10&  -2&  1&  0&  0&  0&  -6&  0&  -4&  0&  4&  0&  -4&  0\\
11&  -5&  1&  0&  3&  0&  -18&  0&  -10&  0&  10&  0&  -10& 0\\
12&  -2&  -1&  0&  0&  -4&  -6&  0&  0&  0&  4&  0&  -4&  0\\
13&  -5&  -1&  0&  -3&  -10&  -12&  0&  0&  0&  10&  0&  -10&  0\\
14&  -1&  0&  0&  0&  0&  -3&  -2&  -3&  1&  0&  -1&  -2&  0\\
15&  -1&  0&  0&  1&  0&  -4&  -2&  -4&  0&  0&  -2&  -2&  0\\
\hline
\end{tabular}\\
\end{center}
\caption{Example of singlet (i.e. no non-Abelian field VEVs) $F$-flat directions to all order for the $SO(10)$ model. Column 1 denotes the flat direction number. Column 2 is the anomalous charge. The remaining columns specify the norm squared VEVs of the respective non-Abelian singlet fields. All other $F$-flat tables are structured in a like manner.}
\label{so10fflat}
\end{table}

\begin{table}[H]
\begin{center}
\begin{tabular}{l|r|rrrrrrrrrrrrrrrrrrrrrrr}
\hline \rule{0pt}{2.6ex}
  & $Q_A$& $\phi_1$& $\psi_1$& $\phi_2$& $\phi_3$& $\psi_2$& $\psi_3$& $\psi_4$& $\psi_5$& $\psi_6$& 
$\psi_7$& $\psi_8$& $\psi_9$& $\phi_4$& $\phi_5$& $\psi_{10}$& $\psi_{11}$& $\psi_{12}$&
$\phi_7$& $\phi_6$& $E^c_1$& $E^c_2$& $E^c_3$\\
\hline
${\cal D}_1$&   0&  1&  0&  0&  0&  0&  0&  0&  0&  0&  0&  0& -2&  0&  0&  0&  2& -2&  0&  0&  0&  0&  0\\
${\cal D}_2$&   0&  0&  0&  0&  0&  0&  0&  0&  1&  0&  0&  0& -1&  0&  0&  0&  1&  0&  0&  0&  0&  0&  0\\
${\cal D}_3$&   0&  0&  0&  0&  0&  0&  0&  0&  0&  0&  0&  0&  0&  1&  0&  0& -1& -1&  0&  0&  0&  0&  0\\
${\cal D}_4$&   0&  0&  0&  0&  0&  0&  0&  0&  0&  0&  0&  1& -1&  0&  0&  0& -1& -1&  0&  0&  0&  0&  0\\
${\cal D}_5$&   0&  0&  0&  0&  0&  0&  0&  0&  0&  0&  0&  0&  0&  0&  0&  1& -1&  0&  0&  0&  0&  0&  0\\
${\cal D}_6$&   0&  0&  0&  0&  0&  0&  0&  0&  0&  0&  0&  0&  0&  0&  1&  0&  2&  0&  0&  0&  0&  0&  0\\
${\cal D}_7$&   0&  0&  0&  0&  0&  0&  0&  0&  0&  0&  1&  0& -1&  0&  0&  0&  1& -1&  0&  0&  0&  0&  0\\
\hline
\end{tabular}\\
\end{center}
\caption{Basis set of non-anomalous U(1) singlet $D$-flat directions for flipped $SU(5)$ model \#1.  Since $Q_A=0$ for all the $D$-flat directions, non-Abelian directions needed to be investigated.}
\label{su51dflatsing}
\end{table}

\begin{table}[H]
\begin{center}
\begin{tabular}{l|r|rrrrrrrrrrrrrrrr}
\hline\rule{0pt}{2.6ex}
& $Q_A$ & $H_1$ & $H_2$ & $H_3$ & $F_1$ & $\bar{F}_1$ & $F_2$ & $\bar{F}_2$ & $F_3$ & $\bar{F}_3$& 
$H_4$& $H_{14}$& $H_{13}$& $H_{15}$& $H_7$& $H_8$& $H_9$\\
\hline
$D_1$&   2&  3& -3& -3&  0&  0&  0&  0&  5&  0&  0&  0&  0&  0&  5&  0&  0\\
$D_2$&   2&  3& -3& -3&  5&  0&  0&  0&  0&  0&  0&  0&  0&  0&  0&  0&  0\\
$D_3$&   2&  3& -3& -3&  0&  0&  5&  0&  0&  0&  0&  0&  0&  0&  0&  5&  0\\
$D_4$&   1&  0&  0& -2&  0&  0&  0&  0&  0&  0&  0&  0&  0&  1&  1&  1&  0\\
$D_5$&   1&  0& -2&  0&  0&  0&  0&  0&  0&  0&  0&  0&  1&  0&  1&  0&  1\\
$D_6$&   2&  2& -4& -4&  0&  0&  0&  0&  0&  0&  1&  0&  0&  0&  3&  2&  2\\
$D_7$& 4& -9& -1& -1&  0&  0&  0&  0&  0&  5&  0&  0&  0&  0&  5&  0&  0\\
$D_8$& -6& -9& 19&  9&  0&  0&  0&  5&  0&  0&  0&  0&  0&  0& -10&  -5&  -10\\
$D_9$&  -1& 0&  2&  2&  0&  0&  0&  0&  0&  0&   0&  1&   0&  0& -2& -1& -1\\
$D_{10}$& -6&  -9&  9& 19&  0&  5&  0&  0&  0&  0&  0&  0&  0& 0& -10& -10& -5\\
\hline
\end{tabular}\\
\end{center}
\caption{Basis set of non-anomalous U(1) non-Abelian $D$-flat directions for flipped $SU(5)$ model \#1.}
\label{su51dflatNA}
\end{table}

\begin{table}[H]
\begin{center}
\begin{tabular}{l|r|rrrrrrrrrrrrrrrrrrrr}
\hline \rule{0pt}{2.6ex}
  & $Q_A$ &  $\psi_{10}$& $\psi_{11}$& $\psi_{12}$& $E^c_1$& $E^c_2$& $E^c_3$& $H_1$& $H_2$& $H_3$& 
$H_5$& $H_6$& $H_{10}$& $H_{11}$& $H_{14}$& $H_{12}$& $H_{13}$& $H_{15}$& $H_7$& $H_8$& $H_9$\\
\hline
$\Delta_1$&   0&  0&  0&  0&  0&  0&  0& -1&  1&  1&  0&  0&  0&  0&  0&  1&  0&  0& -1&  0&  0\\
$\Delta_2$&   0&  0&  0&  0&  0&  0&  0& -1&  1&  1&  0&  0&  0&  1&  0&  0&  0&  0&  0& -1&  0\\
$\Delta_3$&   0&  0&  0&  0&  0&  0&  0& -1&  1&  1&  0&  0&  1&  0&  0&  0&  0&  0&  0&  0& -1\\
$\Delta_4$&   0&  0&  0&  0&  0&  0&  0&  2&  0& -2&  1&  0&  0&  0&  0&  0&  0&  0&  0&  1&  0\\
$\Delta_5$&   0&  0&  0&  0&  0&  0&  0&  2& -2&  0&  0&  1&  0&  0&  0&  0&  0&  0&  0&  0&  1\\
$\Delta_6$&   0&  0&  1& -1&  0&  0&  0&  0&  0&  0&  0&  0&  0&  0&  0&  0&  0&  0&  0&  0&  0\\
$\Delta_7$&   0& -1&  0&  1&  0&  0&  0&  2&  0& -2&  0&  0&  0&  0&  0&  0&  0&  0&  0&  0&  0\\
$\Delta_8$&   0&  0&  0&  1&  0&  0&  0&  0&  1& -1&  0&  0&  0&  0&  0&  0&  0&  0&  0&  0&  0\\
$\Delta_9$&  1& 0&  0&  0&  2& -1& -1&  0&  0&  0&   0&  0&  0&  0&  0&  0&  0&  3&  0&  0&  0\\
$\Delta_{10}$& 1&   0&  0&  0& -1& -1&  2&  0&  0&  0&  0&  0&  0&  0&  3&  0&  0&  0&  0&  0& 0\\
$\Delta_{11}$& 1&   0&  0&  0& -1&  2& -1&  0&  0&  0&  0&  0&  0&  0&  0&  0&  3&  0&  0&  0&  0\\
\hline
\end{tabular}\\
\end{center}
\caption{Basis set of non-anomalous U(1) mixed singlet/non-Abelian $D$-flat directions for flipped $SU(5)$ model \#1.}
\label{su51dflatmix}
\end{table}

\begin{table}[H]
\begin{center}
\begin{tabular}{l|r|rrrrrrrrrrrr}
\hline \rule{0pt}{2.6ex}
FD &$Q_A$& $\psi_9$& $\psi_{10}$& $\psi_{11}$& ${\bar H}_1$& ${\bar H}_2$& ${\bar H}_3$& $H_4$& $H_{10}$& $H_{12}$& 
$H_7$& $H_8$& $H_9$\\
\hline
1&  -14&  -2&  6&  0&  -23&  -1&  -1&  6&  1&  22&  0&  16&  15\\
2&  -1&  -2&  6&  0&  0&  -42&  -42&  15&  0&  18&  31& 34&  34\\
3&  -14&  0&  -3&  -3&  -1&  -13&  -7&  7&  0&  21&  0&  14& 14\\
4&  -15&  0&  -3&  -3&  -1&  -15&  -9&  8&  0&  23&  1&  16& 16\\
5&  -16&  1&  -3&  0&  0&  -12&  -12&  9&  0&  24&  1&  16&  16\\
6&  -1&  0&  33&  -33&  -44&  -68&  -2&  23&  0&  24&  45&  46&  46\\
7&  -1&  0&  30&  -30&  -42&  -60&  0&  21&  0&  24&  39&  -42&  42\\
8&  -1&  0&  9&  -9&  0&  -56&  -38&  19&  0&  20&  37&  38&  38\\
9&  -14&  0&  3&  -3&  -13&  -7&  -1&  7&  0&  21&  0&  14&  14\\
\hline
\end{tabular}\\
\end{center}
\caption{Example of non-anomalous U(1) $F$-flat directions for flipped $SU(5)$ model \#1.  These flat directions required VEVs of non-Abelian fields denoted by $H_i$ and ${\bar H}_i$.}
\label{su51fflat}
\end{table}

\begin{sidewaystable}
\begin{centering}
\begin{tabular}{l|r|rrrrrrrrrrrrrrrrrrrrrrrrrrrrr}
\hline \rule{0pt}{2.6ex}
  & $Q_A$& $\phi_1$& $E^c_1$& $\psi_1$& $\psi_2$& $\psi_3$& $E^c_2$& $E^c_3$& $\psi_4$& $\psi_5$& $\psi_6$& 
$\psi_7$& $\psi_8$& $\psi_9$& $\psi_{10}$& $\psi_{11}$& $\psi_{12}$& $\psi_{13}$& $\psi_{14}$& 
$\psi_{15}$& $\psi_{16}$& $\psi_{17}$& $\psi_{18}$& $\psi_{19}$& $\psi_{20}$& $\psi_{21}$& $\psi_{22}$& $\psi_{23}$& $\psi_{24}$\\
\hline
${\cal D}_1$& -1&  0& -5& -3& -3& -2&  1&  7& -4&  0&  3&  0&  0&  0&  0&  0&  0&  0&  0&  0&  0&  0&  0&  0&  0&  0&  0&  0&  0\\
${\cal D}_2$& -1&  0& -2& -3& -3& -2& -2&  7& -4&  0&  0&  0&  0&  0&  3&  0&  0&  0&  0&  0&  0&  0&  0&  0&  0&  0&  0&  0&  0\\
${\cal D}_3$& -1&  0& -5& -3& -3& -2& -2&  7& -1&  0&  0&  0&  0&  3&  0&  0&  0&  0&  0&  0&  0&  0&  0&  0&  0&  0&  0&  0&  0\\
${\cal D}_4$& -1&  0&  1& -3& -3&  1& -2&  4& -1&  0&  0&  0&  0&  0&  0&  0&  0&  0&  0&  0&  0&  0&  3&  0&  0&  0&  0&  0&  0\\
${\cal D}_5$& -1& -3& -2&  0& -3& -2&  1&  1& -1&  0&  0&  0&  0&  0&  0&  0&  0&  0&  0&  0&  0&  0&  0&  0&  0&  0&  0&  0&  0\\
${\cal D}_6$&  1&  0&  2&  0&  0&  2& -1& -1&  1&  0&  0&  0&  0&  0&  0&  0&  0&  0&  0&  0&  3&  0&  0&  0&  0&  0&  0&  0&  0\\
${\cal D}_7$&  1&  0&  2&  3&  3& -4&  2& -1& -2&  0&  0&  0&  0&  0&  0&  0&  0&  0&  0&  0&  0&  0&  0&  0&  0&  0&  0&  0&  3\\
\hline
\end{tabular}
\caption{Basis set of non-anomalous U(1) singlet $D$-flat directions for flipped $SU(5)$ model \#2. This was all that was necessary to find all-order stringent $F$-flat directions.}
\label{su52dflat}
\end{centering}
\end{sidewaystable}

\begin{table}[h!t!p!]
\begin{center}
\begin{tabular}{l|r|rrrrrrrrrrrrrrrrrr}
\hline \rule{0pt}{2.6ex}
FD &$Q_A$& $\phi_1$& $E_1^c$& $\psi_1$& $\psi_2$& $\psi_3$& $E_2^c$& $E_3^c$& $\psi_{4}$& $\psi_{6}$& 
$\psi_{7}$& $\psi_{9}$& $\psi_{10}$& $\psi_{15}$& $\psi_{19}$& $\psi_{20}$& $\psi_{22}$\\
\hline
1&  -3&  0&  -4&  1& 1&  4&  0&  1& 0&  6&  0&  2&  1 & 0 & 0 & 0 & 10\\
2&  -3&  -6&  -4&  1& -5&  -2&  0&  1& 0&  0&  0&  2&  1 & 0 & 0 & 0 & 4\\
3&  -1&  0&  -8&  1& -3&  -2&  0&  3& 0&  0&  4&  4&  1 & 0 & 2 & 0 & 0\\
4&  -2&  0&  -17&  5& -6&  -1&  0&  0& 4&  0&  11&  9&  1 & 0 & 4 & 0 & 0\\
5&  -1&  0&  -8&  -7& -3&  -2&  0&  11& 0&  0&  0&  4&  1 & 0 & 2 & 4 & 0\\
6&  -2&  -3&  -10&  -3& -6&  0&  0&  8& 0&  1&  0&  1&  1 & 4 & 0 & 0 & 0\\
7&  -4&  -7&  -4&  2& -5&  -1&  0&  0& 0&  1&  0&  2&  2 & 0 & 0 & 0 & 7\\
8&  -2&  -2&  -2&  1& -1&  1&  0&  0& 0&  2&  0&  1&  1 & 0 & 0 & 0 & 5\\
9&  -3&  -4&  -4&  1& 3&  0&  0&  1& 0&  2&  0&  2&  1 & 0 & 0 & 0 & 6\\
10&  -4&  -6&  -5&  1& -4&  0&  0&  0& 1&  2&  0&  3&  1 & 0 & 0 & 0 & 8\\
11&  -3&  -1&  -4&  1& 0&  3&  0&  1& 0&  5&  0&  2&  1 & 0 & 0 & 0 & 9\\
12&  -4&  -2&  -5&  2& 0&  4&  1&  0& 0&  7&  0&  2&  1 & 0 & 0 & 0 & 12\\
13&  -4&  -2&  -5&  2& 0&  4&  1&  0& 0&  7&  0&  2&  1 & 0 & 0 & 0 & 12\\
14&  -4&  -7&  -11&  2& -12&  -1&  0&  0& 0&  1&  7&  2&  2 & 0 & 0 & 0 & 0\\
\hline
\end{tabular}\\
\end{center}
\caption{Example of non-anomalous U(1) singlet stringent $F$-flat directions for flipped $SU(5)$ model \#2.}
\label{su52fflat}
\end{table}

\begin{table}[h!t!p!]
\begin{center}
\begin{tabular}{l|r|rrrrrrrrrrrrrrr}
\hline \rule{0pt}{2.6ex}
 &$Q_A$&   $\phi_1^V$&  $\phi_2^V$& $\phi_3^V$&  $\phi_4^V$& $\phi_5^V$&   $\phi_6^V$&  $E_1^c$& $E_2^c$& $E_3^c$& $\psi_1$& $\psi_2$&  $\psi_3$& $\psi_4$&  $\psi_5$&  $\psi_6$  \\ 
 \hline
${\cal D}_1$&1&  0&  7&  0& -2& -2&  1&  0&  0&  0&  6&  0&  0&  0&  0&  6\\
${\cal D}_2$&1&  0&  4& -3& -2&  1& -2&  0&  0&  0&  0&  0&  6&  0&  0&  6\\
${\cal D}_3$&1&  0&  4& -3&  1& -2& -2&  0&  0&  0&  0&  0&  0&  0&  6&  6\\
${\cal D}_4$&0&  0&  0&  0& -1&  1&  0&  0&  0&  0&  0&  0&  0&  2&  0& -2\\
${\cal D}_5$&   0&  1& -1& -1&  1&  0&  0&  0&  0&  0&  0&  0&  0&  0&  0&  0\\
${\cal D}_6$&  0&  0& -1&  0&  0&  0& -1&  2&  0& -2&  0&  0&  0&  0&  0&  0\\
${\cal D}_7$& 0& 0&  1&  1& -1&  0&  1&  0&  0&  0&  0&  2&  0&  0&  0& -2\\
${\cal D}_8$&0&  0&  0&  0& -1& -1&  0&  0&  2& -2&  0&  0&  0&  0&  0&  0\\
\hline
\end{tabular}\\
\end{center}
\caption{Basis set of non-anomalous U(1) singlet $D$-flat directions for flipped $SU(5)$ model \#3. Since none of the anomalous charges for these directions is negative, non-Abelian directions were investigated.}
\label{su53dflatsing}
\end{table}

\begin{sidewaystable}
\scriptsize
\begin{centering}
\begin{tabular}{l|r|rrrrrrrrrrrrrrrrrrrrrrrrrrrrrrrrrrrr}
\hline \rule{0pt}{2.6ex} 
 & $Q_a$ & $\Psi_{1(4)}$ & $\Psi_{2(3)}$ & $\Psi_{5(8)}$ & $\Psi_{6(7)}$ & $\Psi_{9(10)}$ & $\Psi_{11(12)}$ & $\Psi_{13}$ & $\Psi_{14}$ & $\Psi_{15}$ & $\Psi_{16}$ & $\Psi_{17}$  &$\Psi_{18}$ & $\Psi_{19}$ & $\Psi_{20}$ & $\Psi_{21}$ &  F$_1$ & f$_{1b}$ &  F$_2$ & f$_{2b}$ & F$_3$ & f$_{3b}$ & h$_1$ & h$_2$ & \\
& & & H$_{1(1b)}$ &  H$_{2(3)}$ & H$_{4(5)}$ & H$_{6(7)}$ & H$_8$ & H$_9$ & H$_{10}$ & H$_{11}$ & H$_{12}$ & H$_{13}$ & H$_{14}$ & H$_{15}$ & H$_{16}$ & H$_{17}$ & H$_{18}$ & H$_{19}$ & H$_{20}$& H$_{21}$ & H$_{22}$ & H$_{23}$ & H$_{24}$ 
& H$_{25}$ & H$_{26}$ & H$_{27}$ \\ &&& H$_{28}$ & H$_{29}$ & H$_{30}$ & H$_{31}$ & H$_{32}$ & H$_{33}$ & H$_{34}$ & H$_{35}$ & H$_{36}$ & H$_{37}$ & H$_{38}$ & H$_{39}$ & H$_{40}$ & H$_{41}$ & H$_{42}$ & H$_{43}$ & H$_{44}$ & H$_{45}$ & H$_{46}$ & H$_{47}$ &&&& \rule{0pt}{2.6ex} \\ \shrinkheight{1pt} \hline \hline
$D_{1}$ & -1      &0     &-16     &-15      &11      &-1       &8       &0       &0       &0       &0       &0       &0       &0       &0       &0       &0       &0       &0       &0       &0     
          &0       &0      &15\\&&       &0       &0       &0       &0       &0       &0       &0       &0       &0       &0       &0       &0       &0       &0       &0       &0       &0     
          &0       &0       &0       &0       &0       &0       &0\\&&       &0      &12       &0       &0       &0       &0       &0       &0       &0       &0       &0       &0       &0     
          &0       &0       &0       &0      &12       &0       &0\\     
 $D_{2}$ & -1      &0     &-16     &-15      &11      &14      &-7       &0       &0       &0       &0       &0       &0       &0       &0       &0       &0       &0       &0       &0       &0     
          &0      &15       &0\\
\end{tabular}\\
\end{centering}
\end{sidewaystable}

\begin{sidewaystable}
\scriptsize
\begin{center}
\begin{tabular}{l|r|rrrrrrrrrrrrrrrrrrrrrrrrrrrrrrrrrrrr}
\multicolumn{27}{l}{\small\sl  \ldots continued from previous page \rule{0pt}{2.6ex}} \\
\hline \rule{0pt}{2.6ex}
 & $Q_a$ & $\Psi_{1(4)}$ & $\Psi_{2(3)}$ & $\Psi_{5(8)}$ & $\Psi_{6(7)}$ & $\Psi_{9(10)}$ & $\Psi_{11(12)}$ & $\Psi_{13}$ & $\Psi_{14}$ & $\Psi_{15}$ & $\Psi_{16}$ & $\Psi_{17}$  &$\Psi_{18}$ & $\Psi_{19}$ & $\Psi_{20}$ & $\Psi_{21}$ &  F$_1$ & f$_{1b}$ &  F$_2$ & f$_{2b}$ & F$_3$ & f$_{3b}$ & h$_1$ & h$_2$ & \\
& & & H$_{1(1b)}$ &  H$_{2(3)}$ & H$_{4(5)}$ & H$_{6(7)}$ & H$_8$ & H$_9$ & H$_{10}$ & H$_{11}$ & H$_{12}$ & H$_{13}$ & H$_{14}$ & H$_{15}$ & H$_{16}$ & H$_{17}$ & H$_{18}$ & H$_{19}$ & H$_{20}$& H$_{21}$ & H$_{22}$ & H$_{23}$ & H$_{24}$ 
& H$_{25}$ & H$_{26}$ & H$_{27}$ \\ &&& H$_{28}$ & H$_{29}$ & H$_{30}$ & H$_{31}$ & H$_{32}$ & H$_{33}$ & H$_{34}$ & H$_{35}$ & H$_{36}$ & H$_{37}$ & H$_{38}$ & H$_{39}$ & H$_{40}$ & H$_{41}$ & H$_{42}$ & H$_{43}$ & H$_{44}$ & H$_{45}$ & H$_{46}$ & H$_{47}$ &&&& \rule{0pt}{2.6ex} \\ \shrinkheight{1pt} \hline \hline
$D_{16}$&  1      &0     &-20     &-33      &13       &4      &-2       &0      &12       &0       &0       &0       &0       &0       &0       &0       &0       &0       &0       &0       &0     
          &0       &0       &0\\&&       &0       &0       &0       &0       &0       &0       &0       &0       &0       &0       &0       &0       &0       &0       &0       &0       &0     
          &0       &0       &0       &0       &0       &0       &0\\&&       &0      &24       &0       &0       &0       &0       &0       &0       &0       &0       &0       &0       &0     
          &0       &0       &0       &0      &24       &0       &0\\     
 $D_{17}$&  1      &0      &-4      &-5      &-1       &6       &2       &0       &0       &0       &0       &0       &0       &0       &0       &0       &0       &0      &20       &0       &0     
          &0       &0       &0\\
\end{tabular}\\
\end{center}
\end{sidewaystable}

\begin{sidewaystable}
\scriptsize
\begin{center}
\begin{tabular}{l|r|rrrrrrrrrrrrrrrrrrrrrrrrrrrrrrrrrrrr}
\multicolumn{27}{l}{\small\sl  \ldots continued from previous page \rule{0pt}{2.6ex}} \\
\hline \rule{0pt}{2.6ex}
 & $Q_a$ & $\Psi_{1(4)}$ & $\Psi_{2(3)}$ & $\Psi_{5(8)}$ & $\Psi_{6(7)}$ & $\Psi_{9(10)}$ & $\Psi_{11(12)}$ & $\Psi_{13}$ & $\Psi_{14}$ & $\Psi_{15}$ & $\Psi_{16}$ & $\Psi_{17}$  &$\Psi_{18}$ & $\Psi_{19}$ & $\Psi_{20}$ & $\Psi_{21}$ &  F$_1$ & f$_{1b}$ &  F$_2$ & f$_{2b}$ & F$_3$ & f$_{3b}$ & h$_1$ & h$_2$ & \\
& & & H$_{1(1b)}$ &  H$_{2(3)}$ & H$_{4(5)}$ & H$_{6(7)}$ & H$_8$ & H$_9$ & H$_{10}$ & H$_{11}$ & H$_{12}$ & H$_{13}$ & H$_{14}$ & H$_{15}$ & H$_{16}$ & H$_{17}$ & H$_{18}$ & H$_{19}$ & H$_{20}$& H$_{21}$ & H$_{22}$ & H$_{23}$ & H$_{24}$ 
& H$_{25}$ & H$_{26}$ & H$_{27}$ \\ &&& H$_{28}$ & H$_{29}$ & H$_{30}$ & H$_{31}$ & H$_{32}$ & H$_{33}$ & H$_{34}$ & H$_{35}$ & H$_{36}$ & H$_{37}$ & H$_{38}$ & H$_{39}$ & H$_{40}$ & H$_{41}$ & H$_{42}$ & H$_{43}$ & H$_{44}$ & H$_{45}$ & H$_{46}$ & H$_{47}$ &&&& \rule{0pt}{2.6ex} \\ \shrinkheight{1pt} \hline \hline
$D_{31}$&  0      &0      &-2      &-1       &2       &1       &0       &0       &0       &0       &0       &0       &0       &0       &0       &0       &0       &0       &0       &0       &0     
          &0       &0       &0\\&&       &0       &0       &0       &0       &0       &2       &0       &0       &0       &0       &0       &0       &0       &0       &0       &0       &0     
          &0       &0       &0       &0       &0       &0       &0\\&&       &0       &2       &0       &0       &0       &0       &0       &0       &0       &0       &0       &0       &0     
          &0       &0       &0       &0       &0       &0       &0\\     
 $D_{32}$&  0      &0      &-4      &-5       &3       &0       &0       &0       &0       &0       &0       &0       &0       &0       &0       &0       &0       &0       &0       &0       &0     
          &0       &0       &0\\
\end{tabular}\\
\end{center}
\end{sidewaystable}

\begin{sidewaystable}
\scriptsize
\begin{center}
\begin{tabular}{l|r|rrrrrrrrrrrrrrrrrrrrrrrrrrrrrrrrrrrr}
\multicolumn{27}{l}{\small\sl  \ldots continued from previous page \rule{0pt}{2.6ex}} \\
\hline \rule{0pt}{2.6ex}
 & $Q_a$ & $\Psi_{1(4)}$ & $\Psi_{2(3)}$ & $\Psi_{5(8)}$ & $\Psi_{6(7)}$ & $\Psi_{9(10)}$ & $\Psi_{11(12)}$ & $\Psi_{13}$ & $\Psi_{14}$ & $\Psi_{15}$ & $\Psi_{16}$ & $\Psi_{17}$  &$\Psi_{18}$ & $\Psi_{19}$ & $\Psi_{20}$ & $\Psi_{21}$ &  F$_1$ & f$_{1b}$ &  F$_2$ & f$_{2b}$ & F$_3$ & f$_{3b}$ & h$_1$ & h$_2$ & \\
& & & H$_{1(1b)}$ &  H$_{2(3)}$ & H$_{4(5)}$ & H$_{6(7)}$ & H$_8$ & H$_9$ & H$_{10}$ & H$_{11}$ & H$_{12}$ & H$_{13}$ & H$_{14}$ & H$_{15}$ & H$_{16}$ & H$_{17}$ & H$_{18}$ & H$_{19}$ & H$_{20}$& H$_{21}$ & H$_{22}$ & H$_{23}$ & H$_{24}$ 
& H$_{25}$ & H$_{26}$ & H$_{27}$ \\ &&& H$_{28}$ & H$_{29}$ & H$_{30}$ & H$_{31}$ & H$_{32}$ & H$_{33}$ & H$_{34}$ & H$_{35}$ & H$_{36}$ & H$_{37}$ & H$_{38}$ & H$_{39}$ & H$_{40}$ & H$_{41}$ & H$_{42}$ & H$_{43}$ & H$_{44}$ & H$_{45}$ & H$_{46}$ & H$_{47}$ &&&& \rule{0pt}{2.6ex} \\ \shrinkheight{1pt} \hline \hline
$D_{46}$&  0      &0       &3       &4      &-2       &0      &-1       &0       &0       &0       &0       &0       &0       &0       &0       &0       &0       &0       &0       &0       &0     
          &0       &0       &0\\&&       &0       &0       &0       &0       &0       &0       &0       &0       &0       &0       &0       &0       &0       &0       &0       &0       &0     
          &0       &0       &0       &0       &0       &0       &2\\&&       &0      &-2       &0       &0       &0       &0       &0       &0       &0       &0       &0       &0       &0     
          &0       &0       &0       &0      &-4       &0       &0\\    
 $D_{47}$&  0      &0       &0       &1       &0       &1       &0       &0       &0       &0       &0       &0       &0       &0       &0       &0       &0       &0       &0       &0       &0     
          &0       &0       &0\\
\end{tabular}\\
\end{center}
\caption{Basis set of non-anomalous U(1) non-Abelian $D$-flat directions for flipped $SU(5)$ model \#3. The negative Type I directions listed above are sufficient to produce stringent $F$-flat directions.}
\label{su53dflatNA}
\end{sidewaystable}

\normalsize
\begin{table}[t!]
\begin{center}
\begin{tabular}{l|r|rrrrrrrrrrrr}
\hline \rule{0pt}{2.6ex}
FD &$Q_A$& $\psi_1$& $\psi_2$& $\psi_3$& $\psi_4$& $\psi_5$& $\psi_6$& $H_2$& $H_{7}$& $H_{9}$& 
$H_{12}$& $H_{16}$& $H_{29}$\\
\hline
1&  -1& -4&  0&  2& 2&  2&  -6&   0&  6&  6&  0&  6&  0\\
2&  -1&  0&   1&  3& -2& 0& -6& 10&  6&  0&  6&  0&  4\\
3&  -1& -1&  0&   2&  2& -1& -6&  6&  6&  0&  0&  6&  0\\
\hline
\end{tabular}\\
\end{center}
\caption{Example of non-Abelian all-order stringent $F$-flat directions for flipped $SU(5)$ model \#3.}
\label{su53fflat}
\end{table}
\newpage

\section{Conclusions}\label{conclude}
In answer to the posits made in \cite{cfmt}, we have shown with this sampling of models that reduced Higgs do not correlate to a lack of stringent flat directions.  Indeed, it is apparently possible to a find the whole spectrum of possible types of flatness results, some from singlets only, some from non-Abelian only and some from mixed, that also yield a reduced number of Higgs.  A summary of each of the models' gauge groups, number of Higgs and number and type of all-order stringent flat directions is presented in Table \ref{summary}.  As can be seen from these results, it is quite possible to find all-order stringent flat directions and still preserve symmetries that can break to the SM gauge group. 

Models with flat directions that are all-order stringent flat (or stringent flat beyond 17th order) require an alternative SUSY breaking mechanism.  One way to do this is through hidden sector condensates that form at higher energy above the electroweak scale, which requires hidden sector $SU(n)$ or $SO(2n)$ with $n>3$. While SUSY is broken in the hidden sector at the condensation scale, it can be passed on to MSSM-charged states at the 100 GeV to 1 TeV scale through several alternative processes:  gravitational, shadow charges, and gauge kinetic mixing. In each case, weak coupling between the hidden and observable sectors lowers the scale of SUSY breaking among MSSM charged states by (up to) many orders of magnitude. 

The next step in the analysis for these models will concern shadow charges. Usually, the observable and hidden sectors are identified by their independent gauge groups, of which their respective matter states are representations. That is, observable fields do not carry hidden sector charge and vice-versa. However, there may be some extra $U(1)$ charges, as seen in these models, resulting from the compactified six dimensions that are common to both sectors. These are known as shadow charges and can be carried by fields from both sectors. There may also be shadow states that carry only shadow charges.  Thus, observable and hidden sector states may interact very weakly (at generally very high order) in the superpotential via shadow charge coupling. Hidden sector SUSY breaking from condensates pass to the observable sector through such interactions, but at very suppressed scales.  Results concerning hidden sector condensates, their condensation scales and the shadow charge analysis of these and other models is forthcoming.  

\begin{table}[!h]
\begin{center}
\begin{tabular}{c|c|c|c|cc}
\hline
Observable & Hidden & \# Higgs pairs & \# flat dir. & Singlet $D$-flat & NA $D$--flat\\
\hline
$SO(10)$ & $SU(8)\times SU(2)$& 2 & $\ge 15$ &  yes & --\\
$SU(5) \times U(1)$ & $SU(4)\times SU(10)$& 0 & $\ge 9$&  no & yes\\
$SU(5) \times U(1)$ & $SU(8)$& 2+3 & $\ge 14$ & yes & --\\
$SU(5) \times U(1)$ & $SU(2)^4\times SU(4)$& 2 & $\ge 3$&no &yes\\
\hline
\end{tabular}\\
\end{center}
\caption{Summary of results showing there is no direct correlation between the number of untwisted reduced Higgs and a lack of all-order stringent flat directions.  This table also details whether singlet or non-Abelian $D$-flat directions were needed.}
\label{summary}
\end{table}

\bigskip
\medskip
\leftline{\large\bf Acknowledgments}
\medskip

AEF would like to thank the CERN theory division for hospitality. 
This work was supported by the STFC, by the University of Liverpool 
and by Baylor University.


\def\MPLA#1#2#3{{\it Mod.\ Phys.\ Lett.}\/ {\bf A#1} (#2) #3}
\def\IJMP#1#2#3{{\it Int.\ J.\ Mod.\ Phys.}\/ {\bf A#1} (#2) #3}
\def\IJMPA#1#2#3{{\it Int.\ J.\ Mod.\ Phys.}\/ {\bf A#1} (#2) #3}


\newpage
\appendix
\section{States and Charges of $SO(10)$ and $SU(5)$ Models}

{\def\half{\frac{1}{2}}
\def\mhalf{-\frac{1}{2}}
\def\hfw{$\frac{1}{2}$}
\def\malf{-\frac{1}{2}}
\def\mfw{$-\frac{1}{2}$}
\def\sixth{\frac{1}{6}}
\def\third{\frac{1}{3}}
\def\mthird{-\frac{1}{3}}
\def\mbd{$\frac{2,-1}{3}$}
\def\mtd{$-\frac{1}{3}$}
\def\td{$\frac{1}{3}$}
\def\ttd{$\frac{2}{3}$}
\def\mttd{$-\frac{2}{3}$}
\def\mtwothird{\frac{2}{3}}
\def\mtwothird{-\frac{2}{3}}
\def\pmh{$\pm\half$}
\def\sutc{$SU(3)_C$}
\def\sutl{$SU(2)_L$}
\def\suth{$SU(3)_H$}
\def\sutl{$SU(2)_H$}
\def\sutn{$SU(2)^{'}_H$}
\def\UP#1{$U^{'}_{#1}$}
\def\U#1{$U_{#1}$}
\def\UC{$U_C$}
\def\UL{$U_L$}
\def\Ua{$U_A$}

\def\tb{$\bar{3}$}
\def\tbn{\bar{3}}

\def\T#1{$T_{#1}$}
\def\S#1{$S_{#1}$}
\def\H#1{$H_{#1}$}
\def\UR#1{$U_{#1}$}
\def\R#1{$R_{#1}$}
\def\b#1{$b_{#1}$}
\def\hv#1{$h_{#1}$}

\def\p{\phi}
\def\bp{\bar{\phi}}
\def\s{\psi}
\def\bs{\bar{\psi}}
\def\h{h}
\def\bH{\bar{H}}
\def\H{H}
\def\hb{\bar{h}}
\def\bh{\bar{h}}
\def\L{L}
\def\E{E}
\def\bE{\bar{E}}
\def\Nc{N^c}
\def\Q{Q}
\def\dc{d^c}
\def\uc{u^c}
\def\Hs{H^s}
\def\V{V}
\def\Vs{V^s}

\def\K{K}
\def\bh{\bar{h}}
\def\F{F}
\def\bF{\bar{F}}

\begin{center}
\tablefirsthead{
\hline  
State         & $SO(10)$ &\Ua &\UP{1}&\UP{2}&\UP{3}&\UP{4}&\UP{5}&$SU(8)$&$SU(2)$ \rule{0pt}{2.6ex} \\ \shrinkheight{1pt} \hline }
\tablehead{\hline  \multicolumn{10}{|l|}{\small\sl \ldots $SO(10)$ model continued from previous page \rule{0pt}{2.6ex}}\\  \shrinkheight{1pt} \hline  
State         & $SO(10)$ &\Ua &\UP{1}&\UP{2}&\UP{3}&\UP{4}&\UP{5}&$SU(8)$&$SU(2)$ \rule{0pt}{2.6ex} \\  \shrinkheight{1pt} \hline}
\tabletail{ \hline  \multicolumn{10}{|r|}{\small\sl  $SO(10)$ model continued on next page \ldots  \rule{0pt}{2.6ex}} \\ \shrinkheight{1pt} \hline }
\tablelasttail{\hline}
\bottomcaption{States of the $SO(10)$ model and their gauge charges. The names of the states appear in the first column, with the states' 
various charges appearing in the other columns. All $U(1)$ charges are multiplied by a factor of 4 (and similarly for all other models).}
\begin{supertabular}{|c||c||r|rrrrr||cc||}
\hline\hline
$G_1$&  16&  10&     0&    -2&    -4&     0&    20&  1&  1\\   \shrinkheight{1pt}
$G_2$&  16&   6&     2&     0&     2&     0&   -30&  1&  1\\
$G_3$&  16&   12&    -2&     2&     2&     0&    10& 1&  1\\
$h_1$&  10& -12&    -2&    -2&    -2&     0&   -10&  1&  1\\
$\bh_1$&  10&   12&     2&     2&     2&     0&    10&  1&  1\\
$h_2$&  10&  -12&    -2&     2&    -2&     0&   -10&  1&  1\\
$\bh_2$&  10&  12&     2&    -2&     2&     0&    10&  1&  1\\
\hline
$\Phi_1$&  1&   0&     0&     0&     0&     0&     0&  1&  1\\
$\Phi_2$&  1&   0&     0&     0&     0&     0&     0&  1&  1\\
$\Phi_3$&  1&   0&     0&     0&     0&     0&     0&  1&  1\\
\hline 
\hline 
$\phi_1 \left(\bp_1 \right) \rule{0pt}{2.6ex}$  &  1&   0&     0&    -8&     0&     0&     0&  1&  1\\ 
$\phi_2 \left(\bp_2\right)$&  1&  24&     4&     4&     4&     0&    20&  1&  1\\
$\phi_3 \left(\bp_3\right)$&  1&  24&     4&    -4&     4&     0&    20&  1&  1\\
\hline
\hline
$\psi_1 \left(\bs_1\right)$ \rule{0pt}{2.6ex} &  1&  12&    -6&     2&     2&     0&    10&  1&  1\\
$\psi_2 \left(\bs_2\right)$&  1&   0&     2&    -2&     2&     0&   -70&  1&  1\\
$\psi_3 \left(\bs_3\right)$&  1&   8&    -2&     2&   -10&     0&    30&  1&  1\\
$\psi_4 \left(\bs_4\right)$&  1&   0&     0&     4&     0&     0&     0&  1&  1\\
$\psi_5 \left(\bs_5\right)$&  1&   0&     0&     4&     0&     0&     0&  1&  1\\
$\psi_6 \left(\bs_6\right)$&  1&  24&     4&     0&     4&     0&    20&  1&  1\\
$\psi_7 \left(\bs_7\right)$&  1&  12&    -6&    -2&     2&     0&    10&  1&  1\\
$\psi_8 \left(\bs_8\right)$&  1&   0&     2&     2&     2&     0&   -70&  1&  1\\
$\psi_9 \left(\bs_9\right)$&  1&   8&    -2&    -2&   -10&     0&    30&  1&  1\\
$\psi_{10} \left(\bs_{10}\right)$ \rule{0pt}{2.6ex}&  1&  -20&     2&    -2&     2&    16&   -54&  1&  1\\
$\psi_{11} \left(\bs_{11}\right)$&  1&  -20&     0&     0&     0&    16&    16&  1&  1\\
$\psi_{12} \left(\bs_{12}\right)$&  1&  12&     0&     0&     8&   -16&    24&  1&  1\\
$\psi_{13} \left(\bs_{13}\right)$&  1&  -8&    -4&     0&     4&    16&   -44&  1&  1\\
$\psi_{14} \left(\bs_{14}\right)$&  1& -20&     2&     2&     2&    16&   -54&  1&  1\\
\hline  
\hline   
$H_1 \left(\bH_1\right)$ \rule{0pt}{2.6ex} &  1&  24&     0&     0&     0&     8&    48&  1&  2\\
$H_2 \left(\bH_2\right)$&  1&   0&    -4&     4&    -4&     8&    28&  1&  2\\
$H_3 \left(\bH_3\right)$&  1&   0&    -4&    -4&    -4&     8&    28&  1&  2\\
$H_4 \left(\bH_4\right)$&  1& -12&     2&     2&    -6&     8&    18&  1&  2\\
$H_5 \left(\bH_5\right)$&  1&   0&    -2&     2&    -2&     8&   -42&  1&  2\\
$H_6 \left(\bH_6\right)$&  1&  -8&    -2&     2&     6&     8&    -2&  1&  2\\
$H_7 \left(\bH_7\right)$&  1&   0&    -4&     0&    -4&     8&    28&  1&  2\\
$H_8 \left(\bH_8\right)$&  1& -12&     2&    -2&    -6&     8&    18&  1&  2\\
$H_9 \left(\bH_9\right)$&  1&   0&    -2&    -2&    -2&     8&   -42&  1&  2\\
$H_{10} \left(\bH_{10}\right)$&  1&  -8&    -2&    -2&     6&     8&    -2&  1&  2\\
$H_{11} \left(\bH_{11}\right)$&  1&   0&     0&    -2&    -4&    -8&   -28&  8&  1\\
$H_{12} \left(\bH_{12}\right)$&  1&   2&    -2&     2&     2&    -8&   -38&  8&  1\\
$H_{13} \left(\bH_{13}\right)$&  1&  10&     0&     0&     0&     8&    48&  8&  1\\
\hline
\hline
\end{supertabular}
\label{so10gauge}
\end{center}

\begin{center}
\tablefirsthead{
\hline  
State         &$SU(5)$ &\Ua &\UP{1}&\UP{2}&\UP{3}&\UP{4}&\UP{5}&\UP{6}&$SU(4)$& $SO(10)$  \rule{0pt}{2.6ex} \\ \shrinkheight{1pt} \hline }
\tablehead{\hline  \multicolumn{11}{|l|}{\small\sl \ldots $SU(5)$ model \#1 continued from previous page \rule{0pt}{2.6ex}}\\  \shrinkheight{1pt} \hline  
State       &$SU(5)$ &\Ua &\UP{1}&\UP{2}&\UP{3}&\UP{4}&\UP{5}&\UP{6}&$SU(4)$& $SO(10)$ \rule{0pt}{2.6ex}\\  \shrinkheight{1pt} \hline }
\tabletail{ \hline  \multicolumn{11}{|r|}{\small\sl $SU(5)$ model \#1 continued on next page \ldots  \rule{0pt}{2.6ex}} \\ \shrinkheight{1pt} \hline }
\tablelasttail{\hline}
\bottomcaption{States of flipped $SU(5)$ model \#1 and their charges.}
\begin{supertabular}{|c||c||r|rrrrrr||cc||}
\hline
$F_1$&  10&   6&    -2&    -2&     0&     2&    -6&     2&     1&     1\\ \shrinkheight{1pt}
$F_2$&  10&   6&     2&     0&    -4&    -2&     0&     2&     1&     1\\
$F_3$&  10&   6&     0&     2&     2&     4&     6&     2&     1&     1\\
\hline
${\bar f}_1$ \rule{0pt}{2.6ex}&  ${\bar 5}$&   2&    -2&     2&     0&    -6&    -2&    -6&     1&     1\\
${\bar f}_2$&  ${\bar 5}$&   2&     2&     0&     4&    -2&    -8&    -6&     1&     1\\
${\bar f}_3$& ${\bar 5}$&   2&     0&    -2&     2&    -4&    10&    -6&     1&     1\\
\hline
$E^c_1$&   1&   2&    -2&     2&     0&    -6&    -2&    10&   1&   1\\
$E^c_2$&   1&   2&     2&     0&     4&    -2&    -8&    10&   1&   1\\
$E^c_3$&   1&   2&     0&    -2&     2&    -4&    10&    10&   1&   1\\
\hline\hline
$\Phi_1$&   1&   0&     0&     0&     0&     0&     0&     0&   1&   1\\
$\Phi_2$&   1&   0&     0&     0&     0&     0&     0&     0&   1&   1\\
$\Phi_3$&   1&   0&     0&     0&     0&     0&     0&     0&   1&   1\\
\hline\hline
$\p_1 (\bp_1)$ \rule{0pt}{2.6ex}&   1&  -8&     0&     0&     0&   -16&     8&     0&   1&   1\\
$\p_2 (\bp_2)$&   1&  -8&     0&    -4&     8&    -8&    -4&     0&   1&   1\\
$\p_3 (\bp_3)$&   1&  -8&     0&     4&     8&    -8&    -4&     0&   1&   1\\
$\p_4 (\bp_4)$&   1&   0&     0&    -8&     0&     0&     0&     0&   1&   1\\
$\p_5 (\bp_5)$ \rule{0pt}{2.6ex}&   1&   0&     0&     4&    -8&    -8&    12&     0&   1&   1\\
$\p_6$&   1&   0&     0&    -4&    -8&    -8&    12&     0&   1&   1\\ 
$\p_7$&   1&   0&     0&    -4&     8&     8&   -12&     0&   1&   1\\
\hline\hline
$\psi_1 (\bs_1)$ \rule{0pt}{2.6ex} &   1&  -8&     0&    -2&     4&   -12&     2&     0&   1&   1\\
$\psi_2 (\bs_2)$&   1&  -4&     0&    -2&     4&    -4&    -2&     0&   1&   1\\
$\psi_3 (\bs_3)$&   1&  -4&     0&    -2&     4&    -4&    -2&     0&   1&   1\\
$\psi_4 (\bs_4)$&   1&  -4&     0&     2&    -4&   -12&    10&     0&   1&   1\\
$\psi_5 (\bs_5)$&   1&  -4&     0&     6&     4&    -4&    -2&     0&   1&   1\\
$\psi_6 (\bs_6)$&   1&  -4&     0&     0&     0&    -8&     4&     0&   1&   1\\
$\psi_7 (\bs_7)$&   1&  -4&     0&     0&     0&    8&     4&     0&   1&   1\\
$\psi_8 (\bs_8)$&   1&  -4&     0&    -4&     8&     0&    -8&     0&   1&   1\\
$\psi_9 (\bs_9)$&   1&  -4&     0&     4&     8&     0&    -8&     0&   1&   1\\
$\psi_{10} (\bs_{10})$&   1&   0&     0&    -2&     4&     4&    -6&     0&   1&   1\\
$\psi_{11} (\bs_{11})$&   1&   0&     0&    -2&     4&     4&    -6&     0&   1&   1\\
$\psi_{12} (\bs_{12})$&   1&   0&     0&    -6&    -4&    -4&     6&     0&   1&   1\\
\hline\hline
$H_1 (\bH_1)$ \rule{0pt}{2.6ex}&   1&  -8&     0&    -2&    -1&    -2&     2&    -5&  ${\bar 4}$&   1\\
$H_2 (\bH_2)$&   1&  -4&     0&     0&    -5&     2&     4&    -5&  ${\bar 4}$&   1\\
$H_3 (\bH_3)$&   1&  -4&     0&    -2&    -1&     6&    -2&    -5&  ${\bar 4}$&   1\\
\hline
$H_4$ \rule{0pt}{2.6ex}&   1&   2&     0&    -2&    -3&     6&    10&     5&  ${\bar 4}$ &   1\\
$H_5$&   1&   2&     2&     0&    -1&     8&    -8&     5&  ${\bar 4}$&   1\\
$H_6$&   1&   2&    -2&     2&    -5&     4&    -2&     5&  ${\bar 4}$&   1\\
$H_7$&   1&   6&     0&    -2&    -5&     2&    -6&    -5&   4&   1\\
$H_8$&   1&   6&    -2&     0&     1&     8&     0&    -5&   4&   1\\
$H_9$&   1&   6&     2&     2&    -3&     4&     6&    -5&   4&   1\\
$H_{10}$&   1&   6&     2&     2&     2&    -6&     6&     0&   6&   1\\
$H_{11}$&   1&   6&    -2&     0&     6&    -2&     0&     0&   6&   1\\
$H_{12}$&   1&   6&     0&    -2&     0&    -8&    -6&     0&   6&   1\\
$H_{13}$&   1&  10&    -2&     0&    -2&    -2&     8&     0&   1&  10\\
$H_{14}$&   1&  10&     0&     2&     0&     0&   -10&     0&   1&  10\\
$H_{15}$&   1&  10&     2&    -2&     2&     2&     2&     0&   1&  10\\
\hline
\end{supertabular}
\label{su51gauge1} 
\end{center}

\begin{center}
\tablefirsthead{
\hline  
State&$SU(5)$ &\Ua &\UP{1}&\UP{2}&\UP{3}&\UP{4}&\UP{5}&\UP{6}& \UP{7} &$SU(8)$ \rule{0pt}{2.6ex} \\ \shrinkheight{1pt} \hline }
\tablehead{\hline  \multicolumn{11}{|l|}{\small\sl \ldots $SU(5)$ model \#2 continued from previous page \rule{0pt}{2.6ex}}\\  \shrinkheight{1pt} \hline  
State&$SU(5)$ &\Ua &\UP{1}&\UP{2}&\UP{3}&\UP{4}&\UP{5}&\UP{6}& \UP{7} &$SU(8)$ \rule{0pt}{2.6ex}\\  \shrinkheight{1pt} \hline }
\tabletail{ \hline  \multicolumn{11}{|r|}{\small\sl $SU(5)$ model \#2 continued on next page \ldots  \rule{0pt}{2.6ex}} \\ \shrinkheight{1pt} \hline }
\tablelasttail{\hline}
\bottomcaption{States of flipped $SU(5)$ model \#2 and their charges.}
\begin{supertabular}{|c||c||r|rrrrrrr||c||}
\hline
$\F_1$&  10&    20&     2&    -2&     2&    -2&    -2&   -28&     0&   1\\
$\F_2$&  10&    24&     0&     0&     0&     2&    16&  -104&     0&   1\\
$\F_3$&  10&    20&     2&     2&    -2&    -2&    -2&   -28&     0&   1\\
${\bar f}_1$& ${\bar 5}$&   -20&    -6&    -2&     2&    -2&   -34&  -148&     0&   1\\
${\bar f}_2$&  ${\bar 5}$&   -24&    -8&     0&     0&    -2&   -16&   104&     0&   1\\
${\bar f}_3$&  ${\bar 5}$&   -24&    -6&     2&     2&     2&   -34&    16&     0&   1\\
$\E^c_1$&   1&    60&    10&    -2&     2&    -2&    30&    92&     0&   1\\
${\bar E}^c_1$&   1&   -60&   -10&     2&    -2&     2&   -30&   -92&     0&   1\\
$\E^c_2$&   1&    56&     8&     0&     0&    -2&    48&   344&     0&   1\\
$\E^c_3$&   1&    56&    10&     2&     2&     2&    30&   256&     0&   1\\
\hline
$\h_1 (\bh_1)$ \rule{0pt}{2.6ex}&   5&    -4&    -4&     4&     0&     0&   -36&   -12&     0&   1\\
$\h_2 (\bh_2)$&   5&    12&     0&    -2&    -2&    -2&     8&   -52&    -8&   1\\
$\h_3$&   5&    20&     6&     2&    -2&     2&    34&   148&     0&   1\\
$\h_4$&   5&     8&     0&     2&    -2&     2&     8&   112&     8&   1\\
$\h_5$&   5&     8&     0&    -2&     2&     2&     8&   112&    -8&   1\\
$\h_6$&  ${\bar 5}$&   -24&    -6&    -2&    -2&     2&   -34&    16&     0&   1\\
$\h_7$&  ${\bar 5}$&     8&     4&     0&     0&     2&    36&  -152&     0&   1\\
$\h_8$&  ${\bar 5}$&    32&     4&     0&     0&    -2&    -4&   272&     0&   1\\
\hline
$\Phi_1$&   1&     0&     0&     0&     0&     0&     0&     0&     0&   1\\
$\Phi_2$&   1&     0&     0&     0&     0&     0&     0&     0&     0&   1\\
$\Phi_3$&   1&     0&     0&     0&     0&     0&     0&     0&     0&   1\\
\hline
$\p_1 (\bp_1)$ \rule{0pt}{2.6ex}&   1&    36&    -4&    -4&     0&     0&    -4&   108&     0&   1\\
\hline
$\psi_1 (\bs_1)$ \rule{0pt}{2.6ex}&   1&    44&     4&     2&     2&     2&     4&   220&    -8&   1\\
$\psi_2 (\bs_2)$&   1&     4&     0&     4&     0&     2&     0&  -164&     0&   1\\
$\psi_3 (\bs_3)$&   1&    -4&     0&     4&     0&    -2&     0&   164&     0&   1\\
$\psi_4$&   1&    56&    10&    -2&    -2&     2&    30&   256&     0&   1\\
$\psi_5$&   1&    12&     8&     2&     2&    -2&     8&   -52&     8&   1\\
$\psi_6$&   1&    12&     8&    -2&    -2&    -2&     8&   -52&    -8&   1\\
$\psi_7$&   1&    16&     6&     0&     4&     2&    26&  -128&     8&   1\\
$\psi_8$&   1&    20&     6&     0&     0&    -2&    26&  -292&     8&   1\\
$\psi_9$&   1&    12&     6&     0&     0&    -6&    26&    36&    -8&   1\\
$\psi_{10}$&   1&     8&     6&     0&    -4&    -2&    26&   200&    -8&   1\\
$\psi_{11}$&   1&     8&     0&     4&    -4&    -2&     0&  -328&     0&   1\\
$\psi_{12}$&   1&     0&     0&     4&     4&     6&     0&     0&     0&   1\\
$\psi_{13}$&   1&     0&    -4&     0&     0&    -2&    36&   176&     0&   1\\
$\psi_{14}$&   1&     0&     0&    -4&    -4&     6&     0&     0&     0&   1\\
$\psi_{15}$&   1&     8&     0&    -4&     4&    -2&     0&  -328&     0&   1\\
$\psi_{16}$&   1&    40&    -4&     0&     0&     2&    -4&   -56&     0&   1\\
$\psi_{17}$&   1&   -12&    -4&     2&     2&    -2&   -44&  -124&    -8&   1\\
$\psi_{18}$&   1&   -48&    -4&     2&    -2&     2&    -4&   -56&    -8&   1\\
$\psi_{19}$&   1&   -12&    -4&    -2&    -2&    -2&   -44&  -124&     8&   1\\
$\psi_{20}$&   1&    -8&    -6&     0&     4&     2&   -26&  -200&    -8&   1\\
$\psi_{21}$&   1&    -4&    -6&     0&     0&    -2&   -26&  -364&    -8&   1\\
$\psi_{22}$&   1&   -48&    -4&    -2&     2&     2&    -4&   -56&     8&   1\\
$\psi_{23}$&   1&   -12&    -6&     0&     0&    -6&   -26&   -36&     8&   1\\
$\psi_{24}$&   1&   -16&    -6&     0&    -4&    -2&   -26&   128&     8&   1\\
\hline
$\H_1$&   1&    32&     4&     2&     2&    -2&    24&     8&    -4&   8\\
$\H_2$&   1&    32&     6&     0&     0&     2&     6&   -80&    -4&   8\\
$\H_3$&   1&    16&    -2&    -2&     2&     2&    -2&   136&     4&   8\\
$\H_4$&   1&    20&     0&     0&     0&     2&   -20&  -116&     4&   8\\
$\H_5$&   1&    16&    -2&     2&    -2&     2&    -2&   136&     4&   8\\
$\H_6$&   1&    20&    -2&     2&     2&    -2&    -2&   -28&     4&   8\\
$\H_7$&   1&   -28&    -4&     2&    -2&    -2&   -24&  -172&    -4&   8\\
\hline
\end{supertabular}
\label{su51gauge2} 
\end{center}

\begin{center}
\tablefirsthead{
\hline  
State  &$SU(5)$&\Ua &\UP{1}&\UP{2}&\UP{3}&\UP{4}&\UP{5}&\UP{6}&\UP{7}&$SU(2)^{4}\times{SU(4)}$ \rule{0pt}{2.6ex} \\ \shrinkheight{1pt}
}
\tablehead{\hline  \multicolumn{11}{|l|}{\small\sl \ldots $SU(5)$ model \#3 continued from previous page \rule{0pt}{2.6ex}}\\   \hline
State  &$SU(5)$&\Ua &\UP{1}&\UP{2}&\UP{3}&\UP{4}&\UP{5}&\UP{6}&\UP{7}&$SU(2)^{4}\times{SU(4)}$ \rule{0pt}{2.6ex} \\  \hline}
\tabletail{ \hline  \multicolumn{11}{|r|}{\small\sl $SU(5)$ model \#3 continued on next page \ldots  \rule{0pt}{2.6ex}} \\ \shrinkheight{1pt} \hline }
\tablelasttail{\hline}
\bottomcaption{States of flipped $SU(5)$ model \#3 and their charges.}
\begin{supertabular}{|c||c||r|rrrrrrr||c||}
\hline\hline
$F_1$& 10&  8&  0&  2& -2& 0& -2& -4& -2&  (1, 1, 1, 1, 1)\\ \shrinkheight{1pt}
$F_2$& 10&  6&  0&  2&  2&  4&  0&  2& -6&  (1, 1, 1, 1, 1)\\
$F_3$& 10&  6&  0&  2&  0& -2&  4&  2& 12&  (1, 1, 1, 1, 1)\\
\hline
$\bar{f_1}$ {\rule{0pt}{2.6ex}}&  ${\bar 5}$&  8&  0&  -6&  -2&  0&  -2&  -4&  -2& (1, 1, 1, 1, 1)\\
$\bar{f_2}$&  ${\bar 5}$&  2&  0&  -6&   2&  -4&  0&  -2&  -14& (1, 1, 1, 1, 1)\\
$\bar{f_3}$&  ${\bar 5}$&  2&  0&  -6&  0&   -2&  -4&  6& 4&  (1, 1, 1, 1, 1)\\
\hline
$E^{c}_1$&  1&  8&  0&  10&  -2&  0&  -2&  -4&  -2&  (1, 1, 1, 1, 1)\\
$E^{c}_2$&  1&  2&  0&  10&  2&  -4&  0&  - 2&  -14& (1, 1, 1, 1, 1)\\
$E^{c}_3$&  1&  2&  0&  10&	  0&  -2&  -4&	6&  4& (1, 1, 1, 1, 1)\\
\hline
$h_1 (\bar{h}_1)$ {\rule{0pt}{2.6ex}}&  ${\bar 5}$&  -8&  0&  4& -4&  0&  0&  0&  20&(1, 1, 1, 1, 1)\\
$h_2 (\bar{h}_2)$&  ${\bar 5}$&  -8&  0&  4&  4&  0&  0&  0&  20&  (1, 1, 1, 1, 1)\\
\hline
$\Phi_1$ &  1&  0&  0&  0&  0&  0&  0&  0&  0& (1, 1, 1, 1,1)\\
$\Phi_2$&   1&  0&  0&  0&  0&  0&  0&  0&  0& (1, 1, 1, 1, 1)\\
$\Phi_4$&   1&  0&  0&  0&  0&  0&  0&  0&  0& (1, 1, 1, 1, 1)\\
\hline
$\phi_1 (\bar{\phi}_1)$ {\rule{0pt}{2.6ex}}&  1&  4&  0&  0&  0&  0&  -12&  -4&  8& (1, 1, 1, 1, 1)\\
$\phi_2 (\bar{\phi}_2)$&  1&  12&  0&  0&  0&  0&  4&  -12&  24& (1, 1, 1, 1, 1)\\
$\phi_3 (\bar{\phi}_3)$&  1&  -8&  0&  0&  0&  -8&  -8&  0&  -16& (1, 1, 1, 1, 1)\\
$\phi_4 (\bar{\phi}_4)$&  1&  0&  0&  0&  0&  -8&  8&  -8&  0& (1, 1, 1, 1, 1)\\
$\phi_5 (\bar{\phi}_5)$&  1&  0&  0&  0&  4&  4&  0&  -8&  -36& (1, 1, 1, 1, 1)\\
$\phi_6 (\bar{\phi}_6)$&  1&  0&  0&  0&  -4&  4&  0&  -8&  -36& (1, 1, 1, 1, 1)\\
\hline
$\psi_1$&  1&  4&  8&  0&  2&  -2&  2&  8&  -10& (1, 1, 1, 1, 1)\\
$\psi_2$&  1&  4&  -8&  0&  2&  -2&  2&  8&  -10& (1, 1, 1, 1, 1)\\
$\psi_3$&  1&  6&  8&  0&  -2&  -6&  0&  2&  -6& (1, 1, 1, 1, 1)\\
$\psi_4$&  1&  6&  -8&  0&  -2&  -6&  0&  2&  -6& (1, 1, 1, 1, 1)\\
$\psi_5$&  1&  6&  8&  0&  0&  0&  -4&  2& -24& (1, 1, 1, 1, 1)\\
$\psi_6$&  1&  6&  -8&  0&  0&  0&  -4&  2&  -24& (1, 1, 1, 1, 1)\\
\hline\hline
$H_1 (\bar{H}_1)$ {\rule{0pt}{2.6ex}}&  1&  0&  8&  0&  0&  0&  0&  0&  0& (1, 2, 2, 1, 1)\\
$H_2$&  1&  -2&  4&  5&  -2&  5&  -4&  2&  -4& (1, 1, 2, 1, 1)\\
$H_3$&  1&  2&  -4&  -5&  2&  -5&  4&  -2&  4&(1, 1, 1, 2, 1)\\
$H_4$&  1&  4&  0&  0&  2&  -2&  2&  8&  -10& (1, 1, 1, 1, 6)\\
$H_5$&  1&  -4&  0&  0&  -2&  2&  -2&  -8&  10& (1, 2, 1, 2, 1)\\
$H_6$&  1&  10&  0&  0&  -2&  2&  0&  6&  2& (1, 1, 1, 1, 6)\\
$H_7$&  1&  -10&  0&  0&  2&  -2&  0&  -6&  -2& (2, 1, 2, 1, 1)\\
$H_8$&  1&  -4&  4&  5&  0&  3&  2&  8&  10& (1, 1, 2, 1, 1)\\
$H_9$&  1&  4&  4&  5&  0&  3&  -2&  0&  26& (1, 1, 2, 1, 1)\\
$H_{10}$&  1&  2&  4&  5&  -2&  5&  4&  -2&  4& (1, 1, 2, 1, 1)\\
$H_{11}$&  1&  2&  -4&  5&  -2&  5&  4&  -2&  4& (2, 1, 1, 1, 1)\\
$H_{12}$&  1&  -4&  -4&  5&  2&  1&  6&  0&  -8& (2, 1, 1, 1, 1)\\
$H_{13}$&  1&  4&  4&  5&  2&  1&  2&  -8&  8& (1, 1, 1, 2, 1)\\
$H_{14}$&  1&  -2&  4&  5&  -2&  -3&  4&  -6&  -4& (1, 1, 1, 2, 1)\\
$H_{15}$&  1&  -2&  4&  5&  4&  -1&  0&  2&  14& (1, 1, 1, 2, 1)\\
$H_{16}$&  1&  -2&  -4&  5&  -4&  -1&  0&  2&  14& (2, 1, 1, 1, 1)\\
$H_{17}$&  1&  -4&  -4&  5&  2&  1&  6&  0&  -8& (1, 2, 1, 1, 1)\\
$H_{18}$&  1&  0&  -4&  5&  2&  1&  -6&  -4&  0& (1, 2, 1, 1, 1)\\
$H_{19}$&  1&  0&  -4&  -5&  0&  5&  2&  4&  -18& (1, 1, 2, 1, 1)\\
$H_{20}$&  1&  8&  -4&  -5&  0&  5&  -2&  -4&  -2& (1, 1, 2, 1, 1)\\
$H_{21}$&  1&  6&  4&  -5&  2&  3&  4&  2&  12& (1, 2, 1, 1, 1)\\
$H_{22}$&  1&  2&  4&  -5&  2&  3&  -4&  6&  4& (1, 2, 1, 1, 1)\\
$H_{23}$&  1&  6&  4&  -5&  2&  3&  4&  2&  12& (2, 1, 1, 1, 1)\\
$H_{24}$&  1&  8&  4&  -5&  -2&  -1&  2&  -4&  16& (2, 1, 1, 1, 1)\\
$H_{25}$&  1&  0&  -4&  -5&  -2&  -1&  6&  4&  0& (1, 1, 1, 2, 1)\\
$H_{26}$&  1&  2&  -4&  -5&  4&  1&  0&  -2&  -14& (1, 1, 1, 2, 1)\\
$H_{27}$&  1&  2&  4&  -5&  -4&  1&  0&  -2&  -14&(2, 1, 1, 1, 1)\\
$H_{28}$&  1&  0&  -4&  -5&  -2&  -1&  6&  4&  0& (1, 1, 2, 1, 1)\\
$H_{29}$&  1&  4&  -4&  -5&  -2&  -1&  -6&  0&  8& (1, 1, 2, 1, 1)\\
$H_{30}$&  1&  10&  0&  0&  0&  0&  4&  -2&  -16& (1, 2, 2, 1, 1)\\
$H_{31}$&  1&  6&  0&  0&  0&  0&  -4&  2&  -24& (2, 1, 1, 2, 1)\\
$H_{32}$&  1&  10&  0&  0&  0&  0&  4&  -2&  -16& (1, 1, 1, 1, 6)\\
$H_{33}$&  1&  10&  0&  0&  -2&  2&  0&  6& 2& (2,	1, 1, 2, 1)\\
$H_{34}$&  1&  6&  0&  0&  -2&  -6&  0&  2&  -6& (1, 2, 2, 1, 1)\\
$H_{35}$&  1&  12&  0&  0&  2&  -2&  -2&  0&  6&  (2, 1, 1, 2, 1)\\
$H_{36}$&  1&  12&  0&  0&  2&  -2&  -2&  0&  6& (1, 2, 2, 1, 1)\\
$H_{37}$&  1&  2&  -4&  0&  2&  2&  0&  6&  22& (1, 1, 1, 1, 4)\\
$H_{38}$&  1&  -2&  -4&  0&  -2&  -2&  0&  -6&  -22& (1, 1, 1, 1, 4)\\
$H_{39}$&  1&  10&  0&  0&  -2&  2&  0&  6&  2& (1, 2, 1, 2, 1)\\
$H_{40}$&  1&  4&  4&  0&  -2&  -2&  -2&  0&  26& (1, 1, 1, 1, ${\bar 4}$)\\
$H_{41}$&  1&  4&  4&  0&  2&  2&  -2&  -8&  -10&  (1, 1, 1, 1, ${\bar 4}$)\\
$H_{42}$&  1&  8&  -4&  0&  0&  4&  2&  -4&  16& (1, 1, 1, 1, 4)\\
$H_{43}$&  1&  4&  4&  0&  0&  4&  -6&  0&  8& (1, 1, 1, 1, ${\bar 4}$)\\
$H_{44}$&  1&  4&  -4&  0&  0&  -4&  2& -8&  8& (1, 1, 1, 1, 4)\\
$H_{45}$&  1&  0&  4&  0&  0&  -4&  -6&  -4&  0& (1, 1, 1, 1, ${\bar 4}$)\\
$H_{46}$&  1&  12&  0&  0&  2&  -2& -2&  0&  6& (2, 1, 2, 1, 1)\\
$H_{47}$&  1&  0&  0&  0&  0&  0&  0&  0&  0& (2, 1, 1, 2, 6)\\
\hline\hline
\end{supertabular}
\label{su51gauge3}
\end{center}

\bigskip
\medskip

\newpage

\bibliographystyle{unsrt}

\end{document}